\documentclass{article}

\usepackage{arxiv}
\usepackage[T1]{fontenc}    
\usepackage{hyperref}       
\usepackage{url}            
\usepackage{booktabs}       
\usepackage{amsfonts}       
\usepackage{amsmath}
\usepackage{nicefrac}       
\usepackage{microtype}      
\usepackage{graphicx}
\usepackage[numbers]{natbib}
\usepackage{doi}

\usepackage{amssymb}
\usepackage{float}
\usepackage{outlines}
\usepackage{subcaption}
\usepackage{multirow, booktabs}
\usepackage{graphicx} 
\usepackage{adjustbox} 
\usepackage{ulem}

\usepackage{kotex}

\usepackage{xcolor}

\newcommand{\dummySymb}{v}

\newcommand{\solSymb}{x}
\newcommand{\sol}{\boldsymbol \solSymb}

\newcommand{\solArg}[1]{\sol_{#1}}

\newcommand{\nbig}{N}
\newcommand{\spaceSymb}{s}
\newcommand{\nspacedof}{\nbig_\spaceSymb}

\newcommand{\nsmall}{n}

\newcommand{\dummy}{\boldsymbol {\dummySymb}}
\newcommand{\reddummy}{\hat{\dummy}}
\newcommand{\nbasisspace}{{\nsmall_\spaceSymb}}
\newcommand{\solapprox}{\tilde\sol}
\newcommand{\redsolapprox}{\hat\sol}

\newcommand{\basismatspaceSymb}{\Phi}
\newcommand{\basismatspace}{\boldsymbol{\basismatspaceSymb}}

\newcommand{\snapshotSymb}{X}
\newcommand{\snapshots}{\boldsymbol \snapshotSymb}

\def\Ubold{\boldsymbol{U}}
\def\Vbold{\boldsymbol{V}}
\def\Sigmabold{\boldsymbol{\Sigma}}
\def\ubold{\boldsymbol{u}}

\newcommand{\argmin}[1]{\underset{#1}{\text{argmin}}}
\newcommand{\samplematSymb}{Z}
\newcommand{\samplematNT}{\boldsymbol \samplematSymb}
\newcommand{\samplemat}{\samplematNT^T}

\newcommand{\resSymb}{r}
\newcommand{\nbasisres}{\nsmall_\resSymb}

\title{Evaluation of Thermal Control Based on Spatial Thermal Comfort with Reconstructed Environmental Data}

\author{Youngkyu Kim\thanks{Equal contribution} \\
	Intelligence and Interaction Research Center\\
	Korea Institute of Science and Technology\\
	Seongbuk-gu, Seoul 02792, Republic of Korea \\
	\texttt{youngkyu.kim@kist.re.kr} \\
	\And
	Byounghyun Yoo\footnotemark[1] \\
	Intelligence and Interaction Research Center  \\
    Korea Institute of Science and Technology \\
    AI-Robotics, KIST School\\
    Korea National University of Science and Technology\\
	Seongbuk-gu, Seoul 02792, Republic of Korea\\
	\texttt{yoo@byoo.org} \\
	\And
	Ji Young Yun \\
	School of Architecture and Building Science \\
	Chung-Ang University \\
	Dongjak-gu, Seoul 06974, Republic of Korea\\
	\texttt{yjyyjy5350@cau.ac.kr} \\
	\AND
	Hyeokmin Lee \\
	Intelligence and Interaction Research Center\\
	Korea Institute of Science and Technology\\
	Seongbuk-gu, Seoul 02792, Republic of Korea \\
	\texttt{hyeokminlee@kist.re.kr} \\
	\And
    Sehyeon Park \\
	Intelligence and Interaction Research Center\\
	Korea Institute of Science and Technology\\
	Seongbuk-gu, Seoul 02792, Republic of Korea \\
	\texttt{sehyeon.park@kist.re.kr} \\
	\And
	Jin Woo Moon \\
	School of Architecture and Building Science \\
	Chung-Ang University \\
	Dongjak-gu, Seoul 06974, Republic of Korea\\
	\texttt{gilerbert73@cau.ac.kr} \\
	\AND
	Eun Ji Choi\thanks{Corresponding author} \\
	School of Architecture and Building Science \\
	Chung-Ang University \\
	Dongjak-gu, Seoul 06974, Republic of Korea\\
	\texttt{ejjchl77@cau.ac.kr} \\
}
\date{}

\renewcommand{\shorttitle}{\textit{arXiv} Template}

\begin{document}
\maketitle

\begin{abstract}
Achieving thermal comfort while maintaining energy efficiency is a critical objective in building system control. Conventional thermal comfort models, such as the Predicted Mean Vote (PMV), rely on both environmental and personal variables. However, the use of fixed-location sensors limits the ability to capture spatial variability, which reduces the accuracy of occupant-specific comfort estimation. 
To address this limitation, this study proposes a new PMV estimation method that incorporates spatial environmental data reconstructed using the Gappy Proper Orthogonal Decomposition (Gappy POD) algorithm. In addition, a group PMV-based control framework is developed to account for the thermal comfort of multiple occupants.
The Gappy POD method enables fast and accurate reconstruction of indoor temperature fields from sparse sensor measurements. Using these reconstructed fields and occupant location data, spatially resolved PMV values are calculated. Group-level thermal conditions are then derived through statistical aggregation methods and used to control indoor temperature in a multi-occupant living lab environment.
Experimental results show that the Gappy POD algorithm achieves an average relative error below 3\% in temperature reconstruction. PMV distributions varied by up to 1.26 scale units depending on occupant location. Moreover, thermal satisfaction outcomes varied depending on the group PMV method employed. These findings underscore the importance for adaptive thermal control strategies that incorporate both spatial and individual variability, offering valuable insights for future occupant-centric building operations.
\end{abstract}


\section{Introduction}\label{sec:intro}
To achieve building sustainability, energy efficiency, and occupant comfort, building system control strategies have continued to evolve. In particular, the use of occupant thermal comfort information is considered a key factor in improving both occupant well-being and building energy performance \cite{afroz2022predictive, altomonte2019indoor, magnier2010multiobjective}. With the increasing adoption of intelligent buildings, there is growing interest in quantitatively assessing thermal comfort and integrating it as a core parameter in building operations \cite{jazizadeh2014human-building, zhong2017development, choi2024impact}. As a result, various methods for assessing thermal comfort have been developed over time \cite{fanger1970thermal, gagge1986standard,fanger2002extension, de1998developing,ngarambe2020use}. Thermal comfort is determined by a complex set of factors, including indoor environmental conditions, dynamic personal characteristics, and psychological responses.
Early thermal comfort models primarily focused on environmental variables due to technological limitations. Over time, however, these models have been extended to incorporate personal factors, grounded in human heat balance equation \cite{2020thermal, gagge1986standard}. 

Among thermal comfort models, the Predicted Mean Vote (PMV) has been widely adopted as an international standard. PMV incorporates four environmental parameters—indoor air temperature (T\textsubscript{AIR}), relative humidity (RH), mean radiant temperature (T\textsubscript{MRT}), air velocity (V\textsubscript{AIR}) —as well as two personal variables: metabolic rate (M) and clothing insulation (I\textsubscript{cl}) \cite{fanger1970thermal}. The PMV value is expressed on a 7-point integer scale: $-3$ (cold), $-2$ (cool), $-1$ (slightly cool), $0$ (neutral), $+1$ (slightly warm), $+2$ (warm), and $+3$ (hot). A PMV of $0$ represents thermal neutrality, and a PMV range of $-0.5$ to $+0.5$ is generally considered thermally comfortable for approximately 90\% of people \cite{2020thermal}. 

In earlier applications, the primary limitation of using PMV for building control was the difficulty of obtaining accurate personal variables. Due to technological constraints, real-time measurement of individual factors—specifically metabolic rate and clothing insulation—was not feasible, and fixed or assumed values were typically used in practice \cite{2020thermal, sung2020application}. However, recent advances in image-based artificial intelligence (AI), thermal imaging, and activity recognition algorithms have enabled non-intrusive, real-time estimation of occupants’ metabolic rates and clothing levels \cite{choi2023seasonal, choi2022deep-vision-based, liu2022automatic, YUN2025112217, lee2020assessment, dziedzic2018measurement}. These developments now allow for more accurate and individualized PMV calculations based on real-time personal data.

Another key limitation in thermal comfort estimation lies in the measurement of environmental variables. While real-time environmental data are essential for accurate PMV calculations, discrepancies between sensor locations and actual occupant positions can significantly compromise the reliability of thermal comfort estimation. In typical buildings, sensors are installed at fixed points, and their readings are often assumed to represent the conditions of the entire space. However, previous studies have demonstrated that factors such as system layout, building geometry, and surrounding objects can cause considerable spatial variation in thermal conditions \cite{shan2020integrated, Choi2024gappy}. Consequently, thermal sensation may vary depending on an occupant’s location, and such discrepancies become more pronounced in cases of localized heating or cooling. To improve the accuracy of thermal comfort assessments, it is therefore necessary to incorporate environmental data that reflect the occupant’s actual position \cite{Choi2024gappy, garcia2010review}. 

However, deploying a high density of sensors around occupants in real building environments poses significant economic and technical challenges, making it impractical for application in typical indoor spaces outside of laboratory settings. To overcome these limitations in sensor deployment, recent studies have increasingly applied Digital Twin technologies that can predict and reconstruct environmental data by accounting for the complex physical characteristics of building spaces\cite{brunello2021virtual, yang2022digital, arowoiya2024digital, clausen2021digital}. 

Among Digital Twin-based environmental prediction approaches, physics-based simulations using Computational Fluid Dynamics (CFD) and data-driven models based on Machine Learning (ML) have been most widely employed \cite{molinaro2021embedding, arsiwala2023digital, cheng2024thermal}. CFD enables high-resolution analysis of airflow and temperature distributions within indoor spaces, making it useful for capturing detailed environmental characteristics. However, its application in real-time scenarios and large-scale buildings is limited due to the complexity of model setup, the difficulty of defining boundary conditions, and the high computational cost. ML-based models, on the other hand, are capable of learning and predicting nonlinear and high-dimensional patterns from environmental data. Nevertheless, they typically require large training datasets and extensive hyperparameter tuning to achieve high performance. These constraints limit their applicability across diverse domains and varying environmental conditions.

Therefore, a faster and more efficient data reconstruction method is required for real-time environmental monitoring applicable to building control. To address this, data reconstruction algorithms that estimate full spatial environmental fields from a limited number of sensors have been introduced \cite{liguori2021indoor, zhao2016reconstructing}. A representative technique in this context is the Gappy Proper Orthogonal Decomposition (Gappy POD) algorithm, which has been widely adopted as a prominent approach for reconstructing missing data from spatially sparse measurement (gappy data) \cite{everson1995karhunen, bui2004aerodynamic, willcox2006unsteady}. By extracting dominant patterns through principal component decomposition, Gappy POD effectively infers unmeasured values. Its computational efficiency and stability make it particularly suitable for real-time applications in building environments. Data reconstruction algorithms provide access to spatially non-uniform indoor environmental information, enabling localized assessment of the conditions surrounding each occupant. This spatial resolution offers the potential for more accurate and individualized thermal comfort evaluation.

However, conventional PMV model is primarily designed to evaluate the thermal comfort of individual occupants. As a result, many previous studies that use PMV as a control variable have focused on single-occupant environments. In real-world multi-occupant settings, where a single HVAC system serves a shared space, it is inherently difficult to satisfy the thermal preferences of all occupants simultaneously. This challenge arises largely from individual differences in thermal sensation, which are not adequately captured by traditional PMV models. To improve overall comfort in such environments, it is necessary to derive a representative control variable that reflects the distribution of thermal comfort across occupants. This highlights the need for group thermal comfort-based control strategies tailored to multi-occupant scenarios.

However, studies that incorporate representative thermal comfort metrics into building control for multi-occupant environments remain limited, and a well-established methodology for calculating group-level thermal comfort has yet to be developed. Accordingly, further research is needed to establish approaches that can compute a single representative PMV value from individual occupant data, which can then serve as a control variable. Some prior studies have applied statistical methods to PMV values to reflect the distribution of thermal comfort among multiple occupants, demonstrating the potential of group PMV-based control strategies \cite{choi2024impact, choi2025multi}.

This study aims to improve PMV estimation accuracy through real-time reconstruction of indoor environmental data and to assess the effectiveness of control strategies that consider the actual PMV of multiple occupants. To this end, the Gappy POD algorithm is used to reconstruct spatial indoor temperature fields, enabling the analysis of real-time PMV accuracy and its variation according to occupant location. A group PMV-based control experiment is conducted in a living lab environment with actual occupants, and the resulting thermal comfort levels are quantitatively analyzed according to different control strategies. Through this approach, the study demonstrates the effectiveness of using Gappy POD for real-time PMV estimation and highlights the potential of group PMV-based control strategies to improve thermal comfort in multi-occupant environments.

\section{Methodology}\label{sec:method}
\subsection{Data Reconstruction Algorithm}\label{sec:data recon alg}
The PMV assessment depends on two key inputs: accurately reconstructed environmental conditions at occupant locations and real-time estimates of personal factors. To infer environmental conditions at an occupant’s location, this study employs a data reconstruction algorithm that estimates spatial environmental fields from sparse sensor measurements.

Specifically, the Gappy Proper Orthogonal Decomposition (Gappy POD) method is adopted, a classical technique for reconstructing full-field data from limited observations \cite{everson1995karhunen, bui2004aerodynamic, willcox2006unsteady}. This method relies on a linear subspace derived from a Proper Orthogonal Decomposition (POD) of snapshot data. POD reduces high-dimensional state data into a space with low-dimensional orthonormal basis using singular value decomposition (SVD), and approximates the full solution in this reduced space.

Once the POD basis is computed, Gappy POD reconstructs the full state using only partial measurements at selected spatial locations. The reconstruction is performed by estimating the POD coefficients (also known as generalized coordinates) that minimize the discrepancy between the measured and reconstructed values. Since the POD basis matrix is reused and only matrix-vector operations are needed, the method is computationally efficient and easy to implement.

In this study, Gappy POD is applied to indoor thermal data to reconstruct temperature fields from a limited number of sensors. This enables high-resolution estimation of the indoor thermal environment without the need for dense sensor deployment, forming the basis for the thermal comfort analysis and control strategies described in later sections.

It is demonstrated that Gappy POD achieves accurate spatial field reconstruction even with a small number of sensors, making it a suitable approach for indoor environmental monitoring.

The mathematical formulation of Gappy POD is as follows. The high-dimensional state \(\sol \in \mathbb{R}^{\nspacedof}\) is approximated using the POD basis \(\basismatspace \in \mathbb{R}^{\nspacedof \times \nbasisspace}\) as:
\begin{equation}\label{eq:gappyPODform}
\sol \approx \solapprox = \solArg{ref} + \basismatspace \redsolapprox,
\end{equation}
where \(\solArg{ref}\) is a reference solution and \(\redsolapprox \in \mathbb{R}^{\nbasisspace}\) denotes the generalized coordinate (POD coefficient) in the reduced space, with $\nspacedof \gg \nbasisspace$. The POD basis is obtained by performing SVD on a snapshot matrix $\snapshots$, which contains data collected across various times and parameter conditions:
\begin{equation}
\snapshots = \Ubold \Sigmabold \Vbold^T,
\end{equation}
where \(\Ubold\), \(\Sigmabold\), and \(\Vbold\) represent the left singular vectors, singular values, and right singular vectors, respectively. The first \(\nbasisspace\) columns of \(\Ubold\) are selected to construct the POD basis, yielding \(\basismatspace = [\ubold_1, \dots, \ubold_{\nbasisspace}]\). When only sparse measurements are available, the POD coefficients \(\redsolapprox\) are estimated by solving the following least-squares problem:
\begin{equation}\label{eq:gappyPODmin}
\redsolapprox = \argmin{\reddummy \in \mathbb{R}^{\nbasisspace}} \lVert \samplemat (\sol - \basismatspace \reddummy) \rVert_2^2,
\end{equation}
where \(\samplemat \in \mathbb{R}^{\nbasisres \times \nspacedof}\) is a sampling matrix that extracts the rows corresponding to the measurement locations, \(\nbasisres\) is the number of measurement points, and \(\samplemat \sol\) is the sparse measurement vector. This problem admits a closed-form solution:
\begin{equation}
\redsolapprox = (\samplemat \basismatspace)^\dagger \samplemat \sol,
\end{equation}
where $^\dagger$ denotes the Moore–Penrose inverse of a matrix. The reconstructed full state is then computed using Eq.~\ref{eq:gappyPODform}. Note that \((\samplemat \basismatspace)^\dagger\) is computed once and reused during reconstruction, which contributes to the overall computational efficiency of the method.

\subsection{Thermal Comport Evaluation}\label{sec:thermal_comfort_evaluation}
To estimate personal variables of PMV in real time, this study used the Personal Factor estimation model (PF model) developed based on the prior work of Choi et al. \cite{choi2024impact}. 
The PF model is a computer vision–based AI framework designed to estimate metabolic rate and clothing insulation from indoor image data in real time.
The model integrates multiple neural network architectures—including OpenPose, convolutional neural networks (CNNs), and Deep Neural Networks (DNNs)—to simultaneously infer both metabolic rate and clothing insulation from a single image. A key advantage of the PF model is its ability to estimate individual personal factors for each occupant, even in multi-occupant environments.

The core process of the model consists of the following steps. First, individuals are detected from indoor images. Then, the detected person’s joint coordinates are extracted, and their posture is classified to estimate activity level. Simultaneously, the model identifies the garments worn by the individual in the same image and classifies them to estimate clothing insulation. The model is capable of estimating five levels of metabolic rate and over 34 combinations of clothing ensembles. The trained model achieved an average F1 score of over 0.93, indicating high classification accuracy.

\begin{figure}[htbp]
    \centering
    \includegraphics[scale = 0.7]{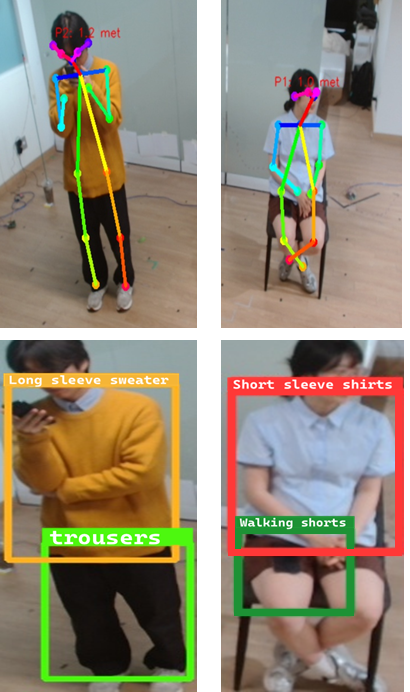}
    \caption{Examples of PF model outputs for estimating personal factors.}
    \label{fig:PFmodel_example}
\end{figure}

Additionally, the applicability of the model was validated in real-world settings, where the mean absolute error (MAE) was 0.04 met for metabolic rate and 0.03 clo for clothing insulation. These results confirm the model's high feasibility for use in actual environments. Accordingly, this study employed the PF model to estimate personal factors for individual occupants. During the experiments, image data were collected at a rate of 2 frames per second (FPS), and personal factors were computed at 20-second intervals. For each interval, the mode value from the captured data was used as the representative estimate. Figure~\ref{fig:PFmodel_example} presents example images illustrating the PF model’s estimation outputs for activity level and clothing insulation used in this study. 

While the PF model enables simultaneous estimation of personal factors for multiple occupants, controlling a shared HVAC system requires a single representative value that reflects the thermal comfort of all occupants. In this study, the conventional PMV metric is modified to derive such a representative value, which is then used as a control variable. To achieve this, three statistical aggregation methods are applied to compute group-level PMV values from individual thermal comfort data \cite{choi2025multi}. 

The first method, PMV\textsubscript{MEDIAN}, uses the median value of all individual PMV data. If the number of data points is odd, the central value is selected; if even, the average of the two middle values is used. This approach is less sensitive to outliers and effectively captures the central tendency of the distribution. 

The second method, PMV\textsubscript{WA}, applies a weighted average technique, in which each individual PMV value is assigned a weight based on its relative importance. The weighted average is computed using Eq.~\ref{eq:pmvwa}, where $n$ denotes the number of individuals (i.e., the number of individual PMV values). Unlike the median-based approach, this method gives greater weight to PMV values associated with thermal discomfort, thereby incorporating the influence of outliers more deliberately. The weighting scheme is defined in Eq.~\ref{eq:pmvwa_weight}. PMV values with absolute magnitudes less than or equal to 1 are considered to indicate relatively comfortable conditions and are assigned a weight of 1. For PMV values outside this range, the square of the PMV is used as the weight. This approach is based on the characteristic that the Predicted Percentage of Dissatisfied (PPD) increases nonlinearly with the square of the PMV, meaning that higher discomfort levels result in greater weights \cite{ashrae55}. 

\begin{equation}\label{eq:pmvwa}
\text{PMV}_\text{WA} = \frac{\sum_{i=1}^{n} w_{i} \cdot \text{PMV}_{i}}{\sum_{i=1}^{n} w_{i}},
\end{equation}
where
\begin{equation}\label{eq:pmvwa_weight}
w_{i} =
\begin{cases}
1, & \text{if } -1.0 \leq \text{PMV}_{i} \leq 1.0 \\
\text{PMV}_i^2, & \text{otherwise}
\end{cases}
\end{equation}

The third method, PMV\textsubscript{MAD}, incorporates the mean absolute deviation (MAD) from the median to compute a representative group PMV. As shown in Eq.~\ref{eq:pmvmad}, the group PMV is calculated by adding the median of individual PMV values to the MAD, which is defined in Eq.~\ref{eq:mad}, where $n$ is the number of individuals (i.e., the number of individual PMV values). This approach is informed by observations that, particularly during the heating season, occupants may still perceive the environment as comfortable even when PMV values are slightly lower than neutral \cite{ashrae55, de2001adaptive}. In this study, the MAD value was added to the PMV median as a correction factor, yielding a higher representative value than the median alone. This adjustment was intended to bridge the gap between objective PMV-based estimates and occupants’ subjective thermal sensations.

\begin{equation}\label{eq:pmvmad}
\text{PMV}_\text{MAD} = \text{PMV}_{\mathrm{median}} + \text{MAD},
\end{equation}
where
\begin{equation}\label{eq:mad}
\quad \text{MAD} = \frac{1}{n} \sum_{i=1}^{n} \left| \text{PMV}_{i} - \text{PMV}_{\mathrm{median}} \right|
\end{equation}

\section{Experiments}
\subsection{Living Lab Set-up}\label{sec:livinglab}
\begin{figure}[htbp]
    \centering
    \includegraphics[width=0.9\linewidth]{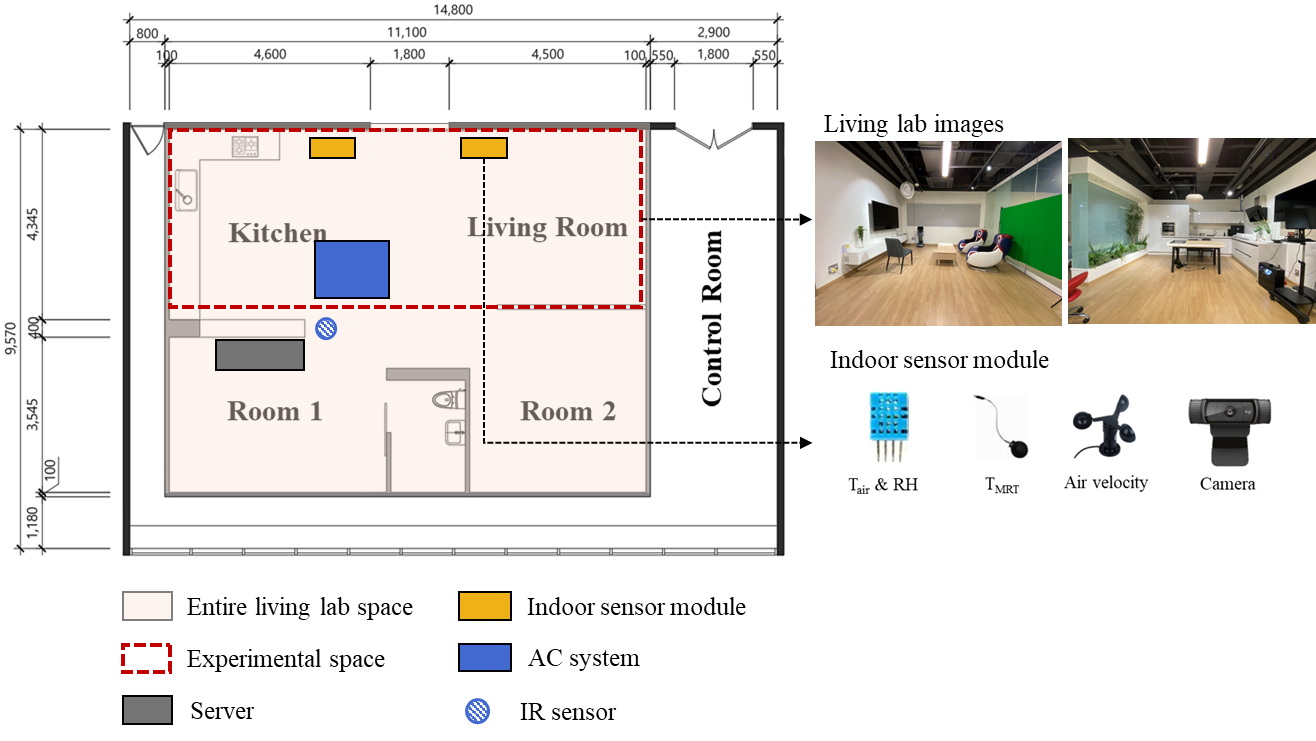}
    \caption{Overview of the living lab environment}
    \label{fig:livinglab_info}
\end{figure}

The living lab is located on the second floor of a research facility in Seoul, South Korea, with a total area of 14.80 m × 9.57 m. As shown in Fig.~\ref{fig:livinglab_info}, the space consists of a control room and a residential environment. The residential area occupies 11.1 m × 8.39 m and includes a living room, kitchen, bedroom, and bathroom. It is designed as an independent space, physically separated from the control room. The openings of the living lab consist of a door and windows connecting to the control room, as well as a main entrance door to the residential area.

The experiment was conducted in the experimental space measuring 11.1 m × 3.96 m, which includes the kitchen and living room areas. The control room was used by the administrator to monitor and manage the experiment. The living lab is equipped with an LRD-N837T inverter-type HVAC system, installed as a four-way ceiling cassette indoor unit. The system is located between the kitchen and living room, with a cooling capacity of 8.3 kW and a heating capacity of 9.3 kW.

In addition, sensor modules were installed in the living lab to collect indoor environmental data and image data. For real-time data acquisition, a microcontroller unit (MCU) consisting of Raspberry Pi and Arduino-based sensors was used. To measure indoor environmental variables—including T\textsubscript{AIR}, RH, T\textsubscript{MRT} and V\textsubscript{AIR}—the sensor modules were mounted on the south wall at a height of 1.2 m, as shown in Fig.~\ref{fig:livinglab_info}, and data were collected at 5-second intervals. 
Details regarding the temperature sensors used for spatial temperature reconstruction are provided in Section \ref{sec:offline}.
To capture indoor images for occupant detection and personal factor estimation, webcams (Logitech C920) were used, recording at 2 FPS. Two cameras were installed at a height of 2 m, positioned to cover the left and right sides of the space. Additionally, an infrared (IR) transmitter was installed near the HVAC system to enable automatic control during the experiment. All data collected from the sensors were transmitted and stored in real time to a local server within the living lab.

\subsection{Data Acquisition}\label{sec:data_acquisition}
\begin{figure*}[!t]
    \centering
    \begin{subfigure}{0.9\textwidth}
        \centering
        \includegraphics[width=\linewidth]{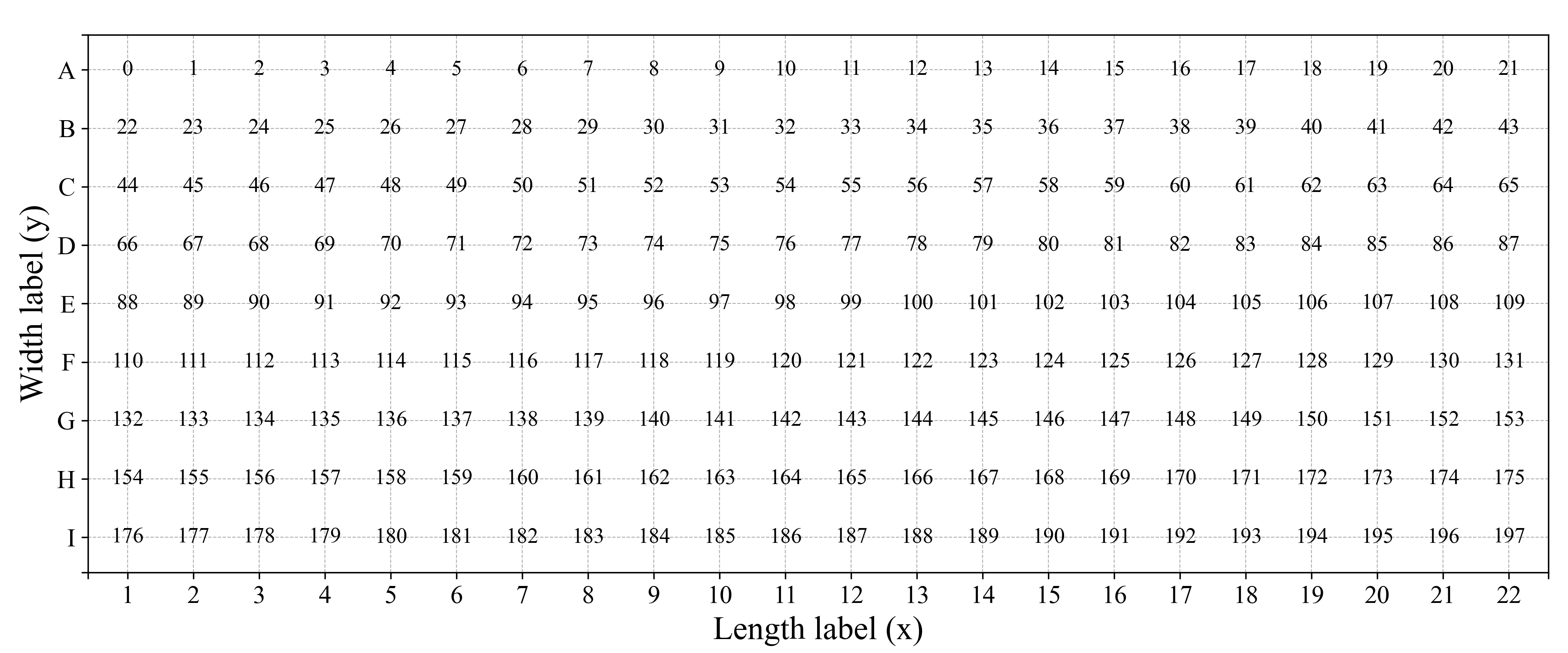}
        \caption{Grid layout (\(9 \times 22\) grid, 198 points). The numbers 0–197 indicate the flat indices corresponding to the tuple of coordinate indices.}
        \label{fig:livinglab_grid}
    \end{subfigure}

    \begin{subfigure}{0.9\textwidth}
        \centering
        \includegraphics[width=\linewidth]{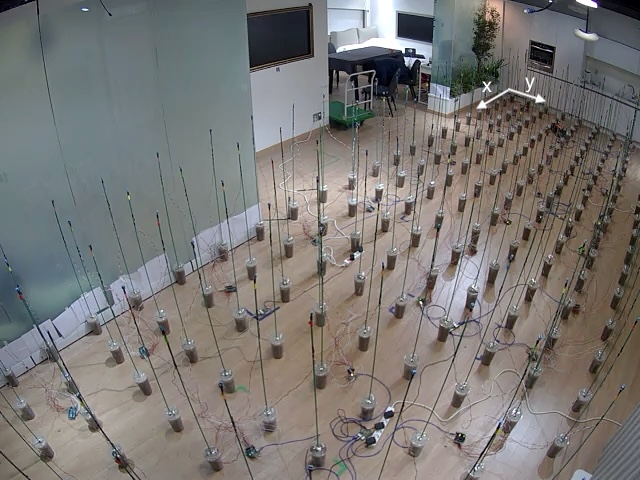}
        \caption{Temperature sensors at grid points}
        \label{fig:livinglab_sensors}
    \end{subfigure}
    \caption{Layout of the living lab}
    \label{fig:livinglab_all}
\end{figure*}

To obtain a training data set (i.e., snapshot data) for constructing the POD basis matrix, the living lab space (9.45 m $\times$ 3.2 m), excluding interior wall thickness, kitchen fixtures, and sensor locations, was divided into a $9 \times 22$ grid. The row and column intervals were set to 40 cm and 45 cm, respectively, resulting in a total of 198 grid points within the rectangular space, as shown in Fig.~\ref{fig:livinglab_grid}. Temperature sensors were installed at a height of 1.2 m at each grid point, as illustrated in Fig.~\ref{fig:livinglab_sensors}. According to the coordinate axes in Fig. \ref{fig:livinglab_grid}, the row direction corresponds to the vertical (y-axis), and the column direction corresponds to the horizontal (x-axis). Each sensor measured temperature at 2-second intervals, and the data were continuously stored in a database.

Snapshot data were collected across the temperature range that included the heating system settings used in the thermal control experiments. Each snapshot scenario consisted of three stages of temperature control: an initial temperature, a first setpoint, and a second setpoint. The specific control sequence was as follows. First, after setting the initial temperature, the heating system operated for 10 minutes, followed by a 10-minute off period. Next, the system was turned on for 15 minutes at the first setpoint temperature, and then switched to the second setpoint for another 15 minutes. Finally, the heating system was turned off and maintained in the off state for an additional 10 minutes. Thus, each scenario lasted for 60 minutes in total. The initial temperature was set to either 22\textdegree C or 23\textdegree C, and the first and second setpoint temperatures were selected from the set \{22\textdegree C, 23\textdegree C, 24\textdegree C, 25\textdegree C, 26\textdegree C\}. This resulted in a total of $2 \times 5 \times 5 = 50$ parameter combinations.

For each scenario, data were collected during the last 2 minutes of three key time points: the first setpoint change, the second setpoint change, and immediately before the final heating system off time. A moving window average with a one-minute interval was applied to each period, resulting in 90 time-series samples per scenario. Consequently, each scenario yielded a two-dimensional snapshot array of size $90 \times 198$.

Among the 50 total scenarios, 8 specific combinations—with initial temperatures of 22\textdegree C or 23\textdegree C, first setpoints of 23\textdegree C or 25\textdegree C, and second setpoints of 23\textdegree C or 25\textdegree C—were used as the validation dataset. The remaining 42 scenarios were used as the snapshot dataset for constructing the POD basis matrix.

\subsection{Experiment Set-up} \label{sec:experiemnt}
The participants consisted of four adults in their twenties—two males and two females. To minimize the effect of body mass index (BMI) on thermal sensation, only individuals within the normal BMI range (18.5 kg/$m^{2}$ < BMI < 24.9 $kg/m^{2}$) were recruited. The average age of the participants was 25.5 years, and the mean BMI was 20.5 kg/$m^{2}$. The experiment was conducted over three days, from January 20 to January 22, 2025, during the heating season. Each day, the experiments were held from 9:00 a.m. to 6:00 p.m., including scheduled breaks.

The thermal environment control experiment was conducted according to four predefined scenarios. As shown in Table~\ref{tab:cases}, personal factors were categorized into two levels for each variable, resulting in a total of four cases. Metabolic rate was classified as either sitting (1.0 met) or standing (1.2 met), while clothing insulation was divided into light clothing (0.5 clo) and heavy clothing (1.0 clo). Table~\ref{tab:scenarios} presents the configuration of the four experimental scenarios (A–D) assigned to each participant. In Scenario A, all participants were assigned different cases, while Scenarios B, C, and D included overlapping cases among some participants. Under identical environmental conditions, the PMV values corresponding to the cases in Table~\ref{tab:cases} increased in the following order: Case 1 < Case 2 < Case 3 < Case 4.

Each of the four scenarios was repeated three times according to the group PMV methods described in Section ~\ref{sec:thermal_comfort_evaluation}, resulting in a total of 12 experimental runs. Subjects 1 through 4 participated from different locations within the space, and temperature data reconstructed for each participant’s specific location was used during the experiment. The occupant locations were assigned based on the living lab grid shown in Fig.~\ref{fig:livinglab_grid}, corresponding to grid positions 114 (row F, column 5), 140 (row G, column 9), 147 (row G, column 16), and 106 (row E, column 19), respectively. For brevity, grid locations are referred to by their row–column indices (e.g., F5 represents the point at row F and column 5) throughout the remainder of this paper, unless coordinate values are explicitly provided.

During the experiment, participants reported their subjective thermal sensation and satisfaction through a structured questionnaire. The survey employed a 7-point Thermal Sensation Vote (TSV) scale, which corresponds to the PMV scale. It followed the point-in-time (short-term) method recommended by ASHRAE 55 \cite{ashrae55}, with data collected every 2 minutes. To ensure accurate responses, participants used a mobile survey platform and avoided actions that could affect thermal sensation, such as eating, drinking, or wearing masks. This study was reviewed and approved by the Institutional Review Board of Chung-Ang University (IRB No. 1041078-20241114-HR-326).

\begin{table}[htbp]
\caption{Personal factor conditions applied for the experiment}
\label{tab:cases}
\centering
\begin{tabular}{@{}lllll@{}}
\toprule
Case                      & 1         & 2         & 3         & 4         \\ \midrule
metabolic rate (met)      & 1.0       & 1.2       & 1.0       & 1.2       \\
clothing insulation (clo) & 0.5       & 0.5       & 1.0       & 1.0       \\ \bottomrule
\end{tabular}
\end{table}

\begin{table}[htbp]
\caption{Subjects scenarios for the experiment}
\label{tab:scenarios}
\centering
\begin{tabular}{@{}lllll@{}}
\toprule
Scenario                  & Subject 1 & Subject 2 & Subject 3 & Subject 4 \\ \midrule
A                         & Case 2    & Case 3    & Case 4    & Case 1    \\
B                         & Case 1    & Case 4    & Case 1    & Case 4    \\
C                         & Case 2    & Case 1    & Case 1    & Case 4    \\
D                         & Case 3    & Case 4    & Case 4    & Case 1    \\ \bottomrule
\end{tabular}
\end{table}

\subsection{Thermal Control Algorithm}\label{sec:algorithm}
\begin{figure}[htbp]
    \centering
    \includegraphics[width=0.9\linewidth]{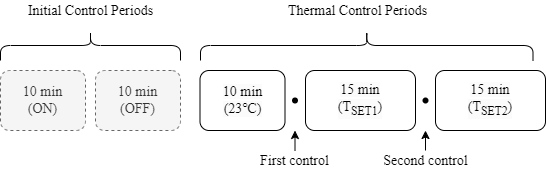}
    \caption{Thermal control sequence}
    \label{fig:control_process}
\end{figure}

The thermal control experiment for each scenario followed the sequence shown in Fig.~\ref{fig:control_process}. To ensure a consistent initial indoor temperature, an initial control period was first applied: the system was turned on at 22°C for 10 minutes and then turned off for the following 10 minutes. Following this step, the thermal control period begins, comprising a preheating phase and two subsequent control points. The preheating phase set the initial temperature to 23 °C and lasts for 10 minutes. The control points refer to the moments when the system adjusts to the designated setpoint temperatures (T\textsubscript{SET1} and T\textsubscript{SET2}). Each control action is performed at 10 and 25 minutes, and the experiment concludes at 40 minutes.

A procedural difference exists between the Gappy POD training data acquisition and the thermal control experiment. While both processes share the same initial control period, the control process includes an additional 10-minute preheating phase at 23\textdegree C prior to the start of the thermal control period.
This difference accounts for real occupancy conditions during the experiments. Unlike the data acquisition phase, which was conducted without occupant movement, participants in the experimental scenarios entered and exited the space during breaks between trials. These door operations introduced cold outdoor air into the room, potentially causing fluctuations in the initial indoor temperature. To minimize this impact and ensure consistent starting conditions, the additional preheating step was incorporated into the experimental procedure.

\begin{figure}[htbp]
    \centering
    \includegraphics[width=0.9\linewidth]{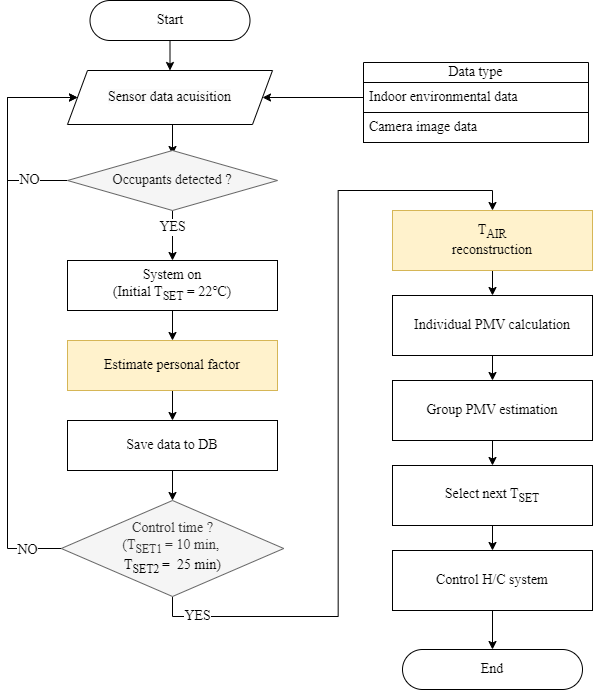}
    \caption{Thermal control algorithm}
    \label{fig:control_algorithm}
\end{figure}

The thermal control algorithm is illustrated in Fig.~\ref{fig:control_algorithm}. To determine the system’s setpoint temperature, the control process iteratively follows the sequence outlined in the figure. First, real-time data are collected from sensors installed throughout the living lab. These include indoor environmental variables as well as indoor image data. Among them, the air temperature data used for the reconstruction algorithm are obtained every 2 seconds from the sensors located at the positions indicated in Fig.~\ref{fig:center_msmt_pts}.
Occupancy detection is performed using the collected image data. Upon detecting occupancy, the HVAC system is activated with a preheating of 23\textdegree C. Subsequently, occupant personal factors—metabolic rate and clothing insulation—are estimated based on the captured images using the personal factor estimation model described in Section \ref{sec:thermal_comfort_evaluation}. All collected data are stored in a centralized database. 

At each control interval, the system evaluates whether an adjustment is necessary. If so, the indoor temperature distribution is reconstructed using 12 temperature measurements taken along the room boundaries. The data reconstruction algorithm estimates zone-specific indoor temperatures based on measurements collected during the two minutes preceding each control point. Using these reconstructed temperature values, along with previously obtained occupant-specific activity levels and clothing insulation data, individual PMV values are calculated. These individual PMVs are then aggregated to estimate a group PMV. The system's optimal setpoint temperature is determined such that the group PMV approaches zero, and this setpoint is subsequently used to control the HVAC system.

During the heating experiments in the living lab, several physical constraints limited the range of feasible system setpoint temperatures. The experimental space was located within the interior of a building, which reduced exposure to outdoor temperature fluctuations and made it difficult to achieve indoor temperatures below 22\textdegree C. Additionally, pre-tests revealed that due to aging HVAC equipment and decreased building airtightness, the system was unable to raise the indoor temperature above 26\textdegree C. Considering these limitations, the operational setpoint temperature range was restricted to between 22\textdegree C and 26\textdegree C throughout the experiment.

\section{Results and Discussion}\label{sec:result}
\subsection{Data Reconstruction Performance}
\subsubsection{Offline Phase}\label{sec:offline}
The number of POD modes used to construct the POD basis matrix should be greater than the intrinsic dimension of the snapshot data, and at the same time, should be less than half the number of measurements to ensure numerical stability in Gappy POD. The former condition ensures that the key features of the data are adequately captured, while the latter helps avoid numerical instability caused by a large condition number during the computation of POD coefficients.

\begin{figure}[htbp]
    \centering
    \includegraphics[width=0.9\linewidth]{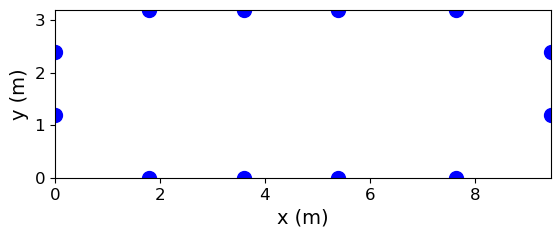}
    \caption{12 measurement points (boundary-based locations)}
    \label{fig:center_msmt_pts}
\end{figure}

In this study, the intrinsic dimension of the snapshot data is four, consisting of three temperature setting parameters and one temporal domain and the number of available measurement points is 12, therefore, the number of POD modes is set to be 5. To evaluate reconstruction performance, we selected 12 sparse measurement locations evenly distributed along the boundaries of the living lab, as shown in Fig.~\ref{fig:center_msmt_pts}.

The reconstruction error was computed using the following definition:
\begin{equation}\label{eq:recon_error}
    \text{reconstruction error}(\%) = \frac{\lVert \sol - \solapprox \rVert_2^2}{\lVert \sol \rVert_2^2} \times 100
\end{equation}
where $\sol$ represents the ground truth temperature field, and $\solapprox$ denotes the reconstructed result.

\begin{figure}[htbp]
    \centering
    \includegraphics[width=0.9\linewidth]{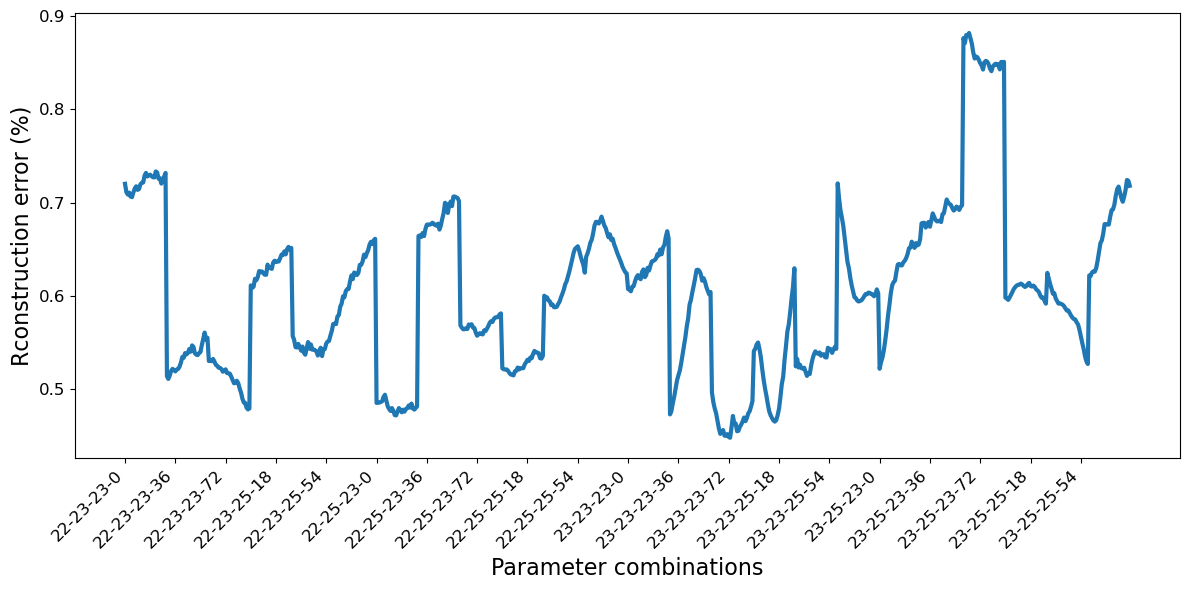}
    \caption{Reconstruction error on the validation dataset using Gappy POD}
    \label{fig:gappy_pod_val}
\end{figure}

The reconstruction performance on the validation dataset is presented in Fig.~\ref{fig:gappy_pod_val}. The x-axis labels are formatted as \texttt{i-j-k-t}, where \texttt{i} is the initial temperature (°C), \texttt{j} and \texttt{k} represent the first and second setpoint temperatures (°C), respectively, and \texttt{t} indicates the index of the moving average window applied to the final 2 minutes of each control point. The average reconstruction error was approximately 0.603\%, with a minimum of 0.448\% and a maximum of 0.882\%.

\begin{figure}[htbp]
    \centering
    \begin{subfigure}{0.9\textwidth}
        \centering
        \includegraphics[width=\linewidth]{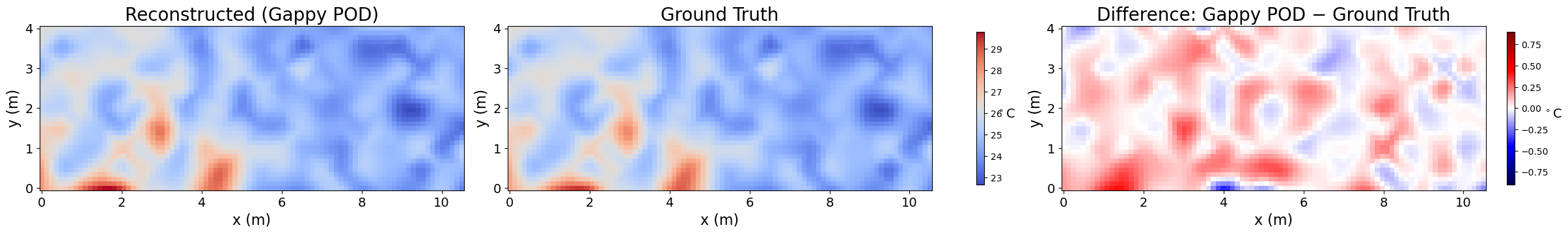}
        \caption{The best case with the lowest reconstruction error.}
        \label{fig:best}
    \end{subfigure}
    \vfill
    \begin{subfigure}{0.9\textwidth}
        \centering
        \includegraphics[width=\linewidth]{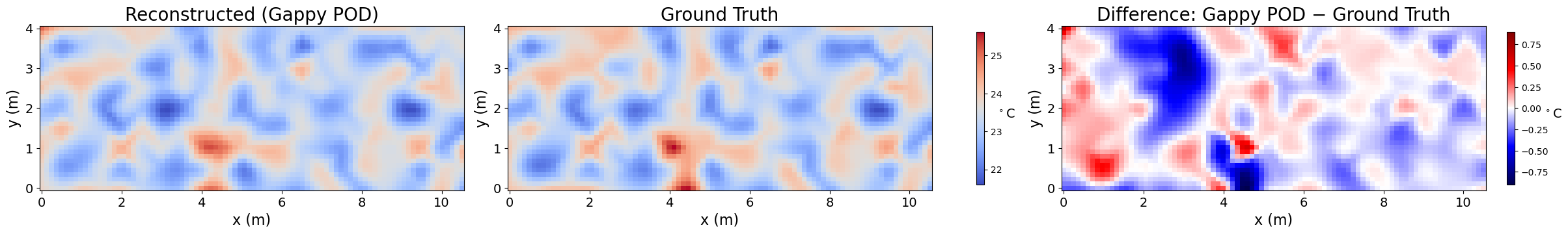}
        \caption{The worst case with the highest reconstruction error.}
        \label{worst}
    \end{subfigure}
    \caption{Comparison of the best and worst reconstruction cases in the validation dataset}
    \label{fig:gappy_pod_val_plt}
\end{figure}

The reconstruction results for the best and worst cases, along with the differences from the ground truth, are shown in Fig.~\ref{fig:gappy_pod_val_plt}. Each column presents the reconstructed temperature field, the ground truth, and their difference. All fields are smoothed using cubic interpolation on a 4$\times$ finer grid for visual clarity.

Even in the worst case, the temperature difference remained below 1\textdegree C. This is because the validation dataset falls within the interpolation range of the parameter sweep used to generate the snapshot (training) data.

\subsubsection{Online Phase}\label{sec:online}
\begin{figure}[htbp]
    \centering
    \includegraphics[width=0.9\linewidth]{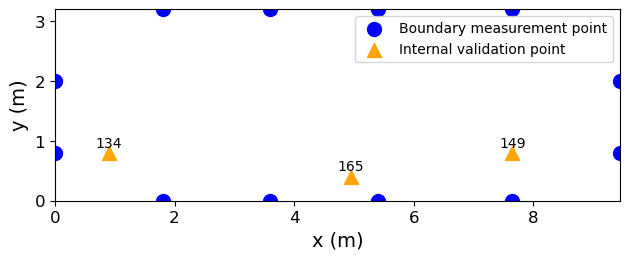}
    \caption{Locations of measurement (boundary) and validation (internal) points}
    \label{fig:msmt_gt_pts}
\end{figure}

During the actual PMV control experiments, the environmental conditions differed from those of the data acquisition phase. Factors such as the presence of occupants, a reduced number of sensors, and varying door-opening frequency in the living lab may have negatively affected reconstruction accuracy. Since it was not feasible to measure the entire temperature field during the experiment, reconstruction performance was evaluated using three additional internal measurement points, in addition to the original 12 boundary sensors. These measurement locations are shown in Fig.~\ref{fig:msmt_gt_pts}, where the boundary points are marked by blue circles and the internal validation points by orange triangles.The numbers above the orange markers denote the flat indices of the corresponding grid points.

\begin{figure}[htbp]
    \centering
    \includegraphics[width=0.9\linewidth]{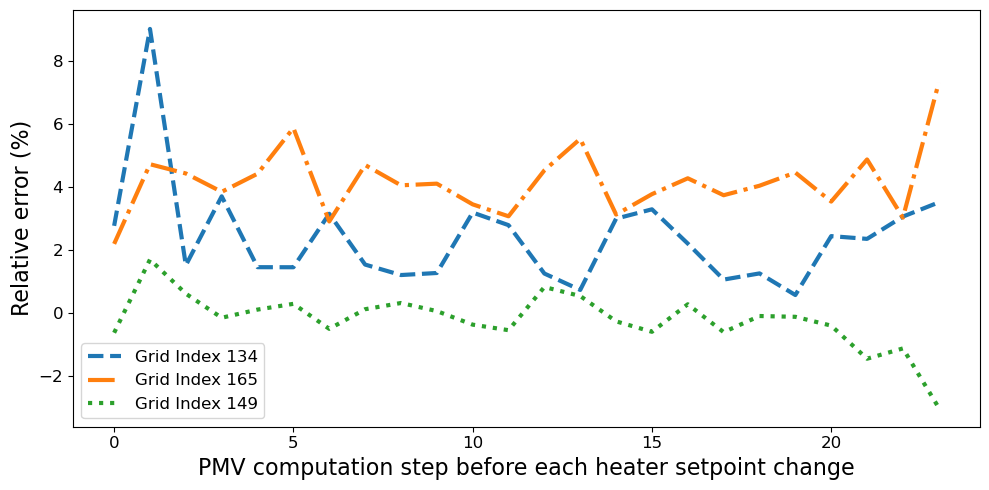}
    \caption{Relative error between reconstructed and measured temperatures}
    \label{fig:gappy_pod_real_plt}
\end{figure}

At each PMV control point during the experiment, the relative error between the reconstructed temperature and the ground truth measured at the internal validation points was computed. The error was computed as $(T_{\text{measured}}-T_{\text{reconstructed}})/ T_{\text{measured}} \times 100$. The results are plotted in Fig.~\ref{fig:gappy_pod_real_plt}.

The average relative error is 2.13\%, corresponding to a temperature difference of 0.56\textdegree C. 

\subsection{Thermal Comfort Analysis}

\subsubsection{Results of Spatial PMV Estimation}\label{sec:PMV_variability}
\begin{figure}[htbp]
    \centering
    \begin{subfigure}{0.9\textwidth}
        \centering
        \includegraphics[width=\linewidth]{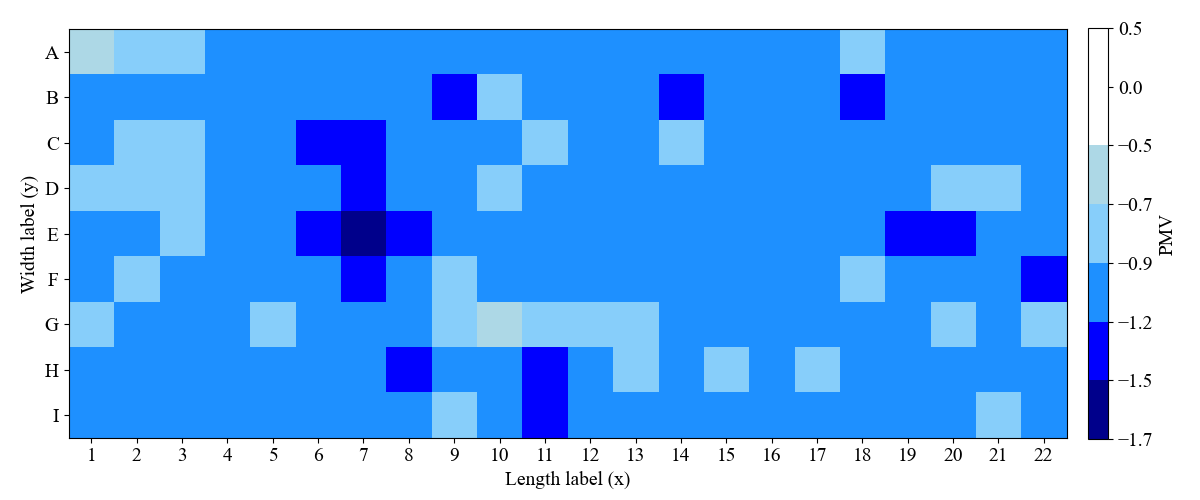}
        \caption{After 10 minutes of heating with a setpoint of 23$^{\circ}$C}
        \label{fig:discomfort_reconstruct_23}
    \end{subfigure}
    \vfill
    \begin{subfigure}{0.9\textwidth}
        \centering
        \includegraphics[width=\linewidth]{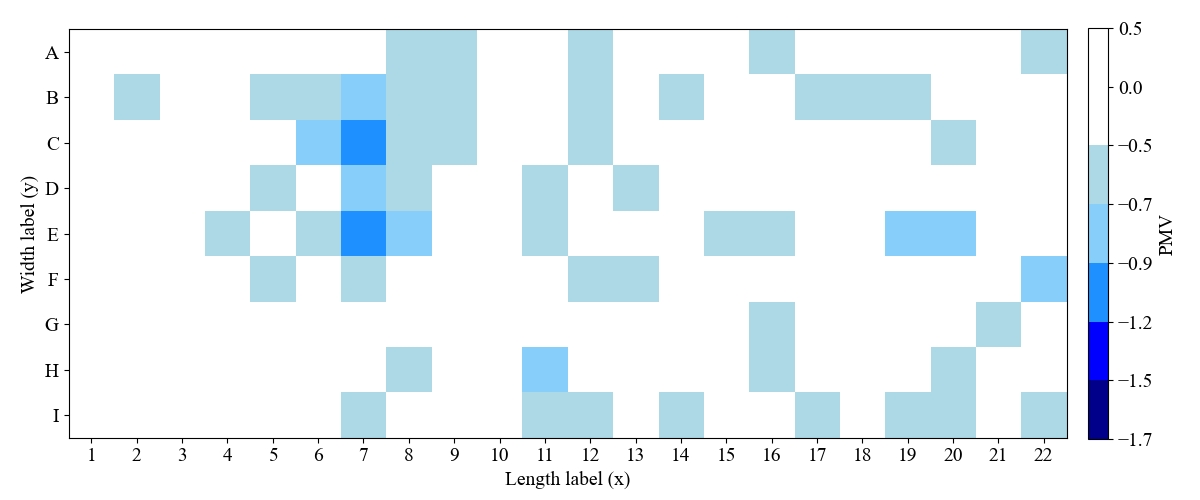}
        \caption{After 15 minutes of heating with a setpoint of 26$^{\circ}$C}
        \label{fig:discomfort_reconstruct_26}
    \end{subfigure}        
    \caption{Spatial distribution of PMV based on reconstructed indoor temperature fields}
    \label{fig:discomfort_reconstruct}
\end{figure}

Fig.~\ref{fig:discomfort_reconstruct} presents the spatial distribution of PMV values calculated based on the reconstructed indoor temperature data within the living lab space. Specifically, PMV was estimated using the reconstructed temperature field. Temperature data for computing PMV as shown in Fig.~\ref{fig:discomfort_reconstruct_23} and Fig.~\ref{fig:discomfort_reconstruct_26} are reconstructed after 10 minutes of heating with a setpoint temperature of 23$^{\circ}$C and 15 minutes of heating with a setpoint temperature of 26$^{\circ}$C, respectively. The remaining environmental variables required for PMV calculation were measured during the experiment, while personal factors were assumed to be 1.2~met and 0.5~clo. In these subfigures, thermally comfortable regions (-0.5 < PMV < +0.5) are indicated in white, while thermally uncomfortable zones (PMV < -0.5) are shown in increasingly darker shades of blue as the magnitude of discomfort (i.e., |PMV|) increases.

In Fig.\ref{fig:discomfort_reconstruct_23}, PMV values range from -1.59 to -0.62, indicating a predominantly thermally uncomfortable environment under the 23$^{\circ}$C setpoint condition. In contrast, Fig.\ref{fig:discomfort_reconstruct_26} shows a wider range from -1.06 to 0.20, reflecting a general shift toward thermal comfort range. However, the PMV variation of up to 1.26 across the space highlights the strong influence of occupant location. These findings underscore the importance of accounting for spatial temperature distribution when estimating PMV in multi-occupant environments.

The heating system in the living lab is located near the bottom-left corner of the space, corresponding to the area around grid points H7, H8, I7, and I8. This positioning results in uneven airflow distribution and limited diffusion, potentially causing residual cool zones on the opposite side of the room. As a consequence, noticeable differences in PMV values were observed even near the HVAC unit. These findings confirm the presence of spatial thermal non-uniformity, which can be effectively quantified using the data reconstruction algorithm. Thermal field reconstruction contributes to a more precise assessment of PMV distribution throughout the space.

\begin{figure}[htbp]
    \centering
    \includegraphics[width=0.9\linewidth]{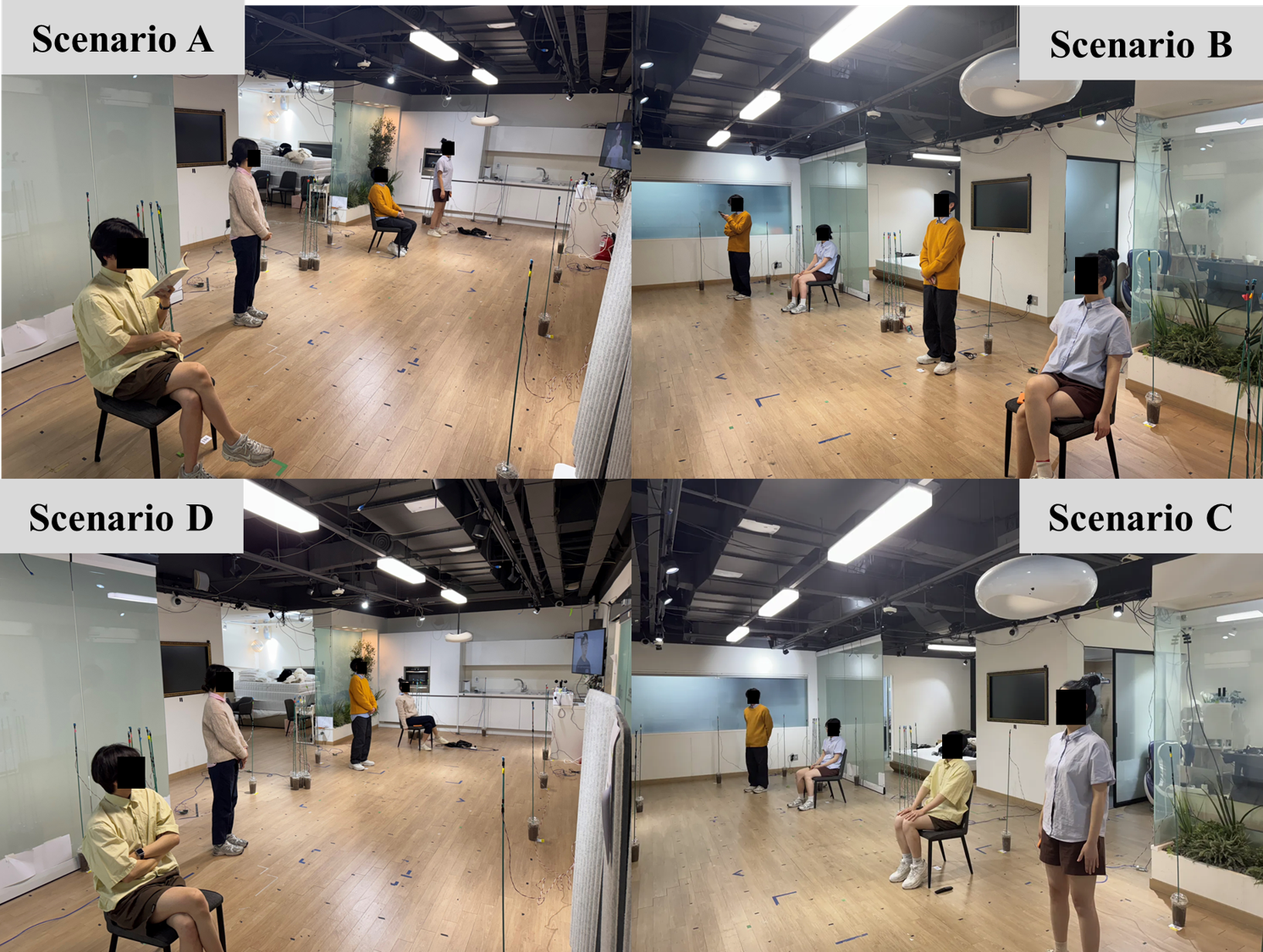}
    \caption{Experimental images for each scenario}
    \label{fig:experiment_imgs}
\end{figure}

Additionally, thermal control experiments were conducted for occupants located at positions F5, G9, G16, and E19 in the living lab, based on the scenario-specific personal factor conditions described in Table~\ref{tab:scenarios}. During the experiments, participants adjusted their activity levels and clothing types according to each scenario. Representative images illustrating these experimental conditions are provided in Fig.~\ref{fig:experiment_imgs}.

The estimation accuracy of personal factor for each occupant was evaluated using the mean absolute error (MAE) during the experimental procedure. The PF model estimates of metabolic rate and clothing insulation from image data at 2 FPS, storing the most frequent value every 20 seconds. As a result, three data points per minute were obtained. Over the total experiment duration (40 minutes per scenario × 4 scenarios × 3 control points = 480 minutes), a total of 1,440 personal factor data points were collected per participant. In this study, data from four participants yielded 5,652 personal factor entries in total. Some missing values were observed due to sensor noise and communication errors. The overall MAE between the estimated and ground truth values was 0.011 met for metabolic rate and 0.029 clo for clothing insulation, consistent with the performance level reported in previous studies on the PF model \cite{choi2024impact}.

Based on the estimated personal factors and the reconstructed indoor temperature data obtained during the experiment, average PMV values were compared for each participant, as summarized in Table~\ref{tab:PMV_difference}. These values represent the mean PMV computed across three repeated experimental runs and were evaluated at the end of the 10-minute preheating phase. Notable differences in PMV were observed depending on both personal factors and occupant location, even under the same controlled temperature condition of 23$^\circ$C during that period.

In Scenario A, all participants had different activity and clothing conditions, resulting in distinct PMV values for each individual. In Scenarios C and D, subjects 2 and 3 shared identical personal factor conditions. Specifically, in Scenario C, both subjects performed Case 1, which involved low activity and light clothing; however, their average PMV values differed by approximately 0.68 (-0.78 for subject 2 and -1.46 for subject 3). In Scenario D, where both subjects performed Case 4 with high activity and heavy clothing, the PMV difference was approximately 0.32. This outcome is consistent with the spatial distribution of PMV shown in Fig.~\ref{fig:discomfort_reconstruct}, where Subject 2 (G9) exhibited a higher PMV than Subject 3 (G16), despite identical personal conditions. This discrepancy is attributed to differences in local thermal environments.

These results demonstrate that incorporating reconstructed temperature data and individual personal factors allows for precise PMV estimation at each occupant’s location. The thermal control results based on these estimations are presented in the following section.

\begin{table}[htbp]
\caption{PMV values calculated after 10 minutes of a 23$^\circ$C preheating phase, averaged across three repeated experiments}
\label{tab:PMV_difference}
\centering
\begin{tabular}{@{}lllll@{}}
\toprule
Scenario & Subject 1 & Subject 2 & Subject 3 & Subject 4 \\ \midrule
A        & -1.13     & 0.05      & -0.07     & -1.88     \\
B        & -1.44     & 0.31      & -1.37     & -0.11     \\
C        & -0.99     & -0.78     & -1.46     & -0.03     \\
D        & -0.48     & 0.16      & -0.16     & -2.02     \\ \bottomrule
\end{tabular}
\end{table}

\subsubsection{Thermal Control Results Based on Group PMV}\label{sec:TC_results}
\begin{figure}[htbp]
    \centering
    \begin{subfigure}{0.45\textwidth}
        \centering
        \includegraphics[width=\linewidth]{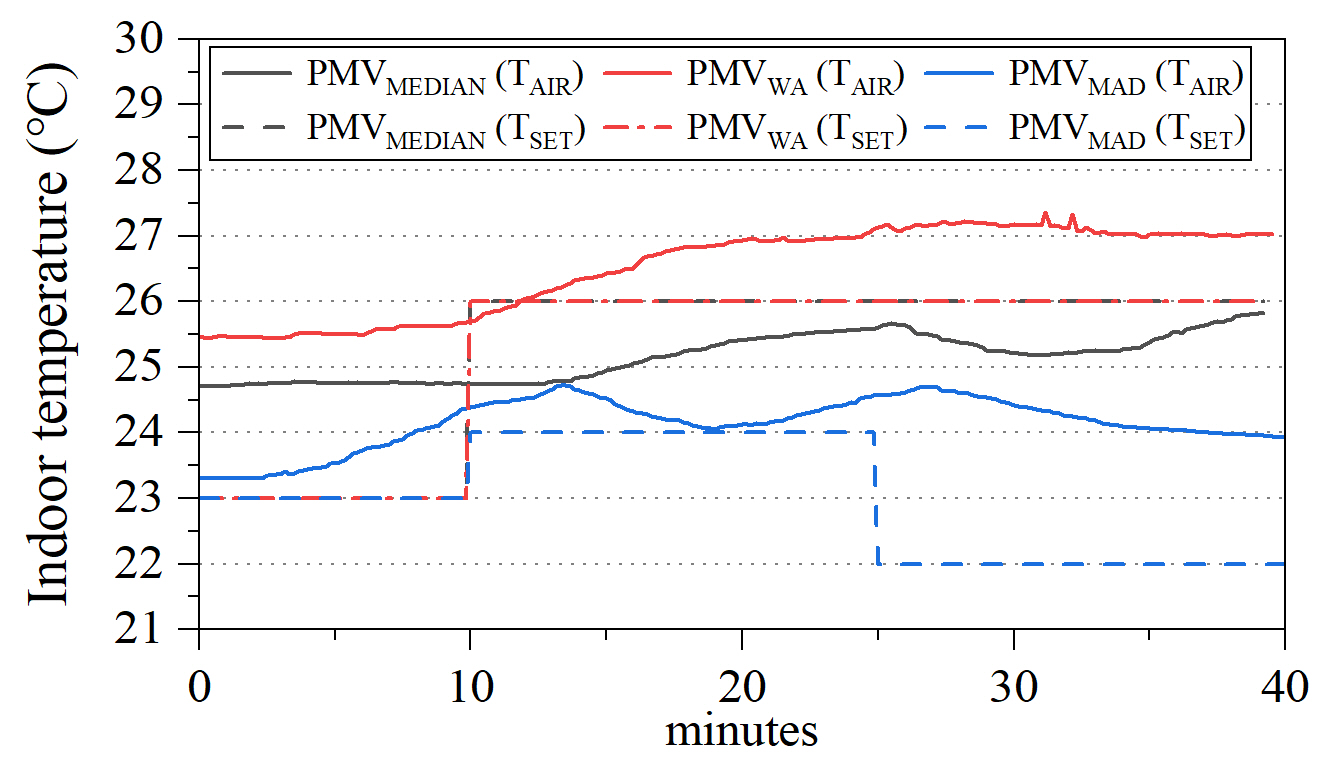}
        \caption{Scenario A}
        \label{fig:results_temp_a}
    \end{subfigure}
    \hfill
    \begin{subfigure}{0.45\textwidth}
        \centering
        \includegraphics[width=\linewidth]{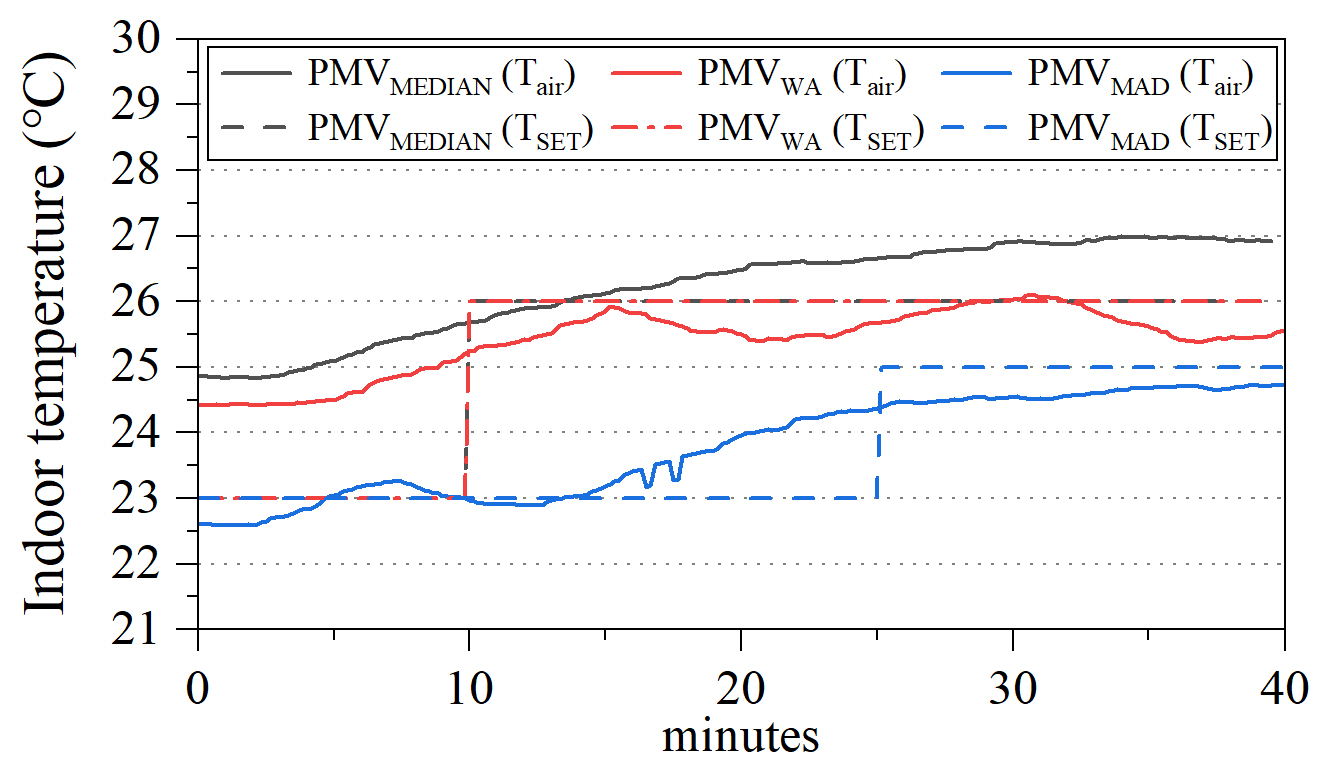}
        \caption{Scenario B}
        \label{fig:results_temp_b}
    \end{subfigure}
    \vfill
    \begin{subfigure}{0.45\textwidth}
        \centering
        \includegraphics[width=\linewidth]{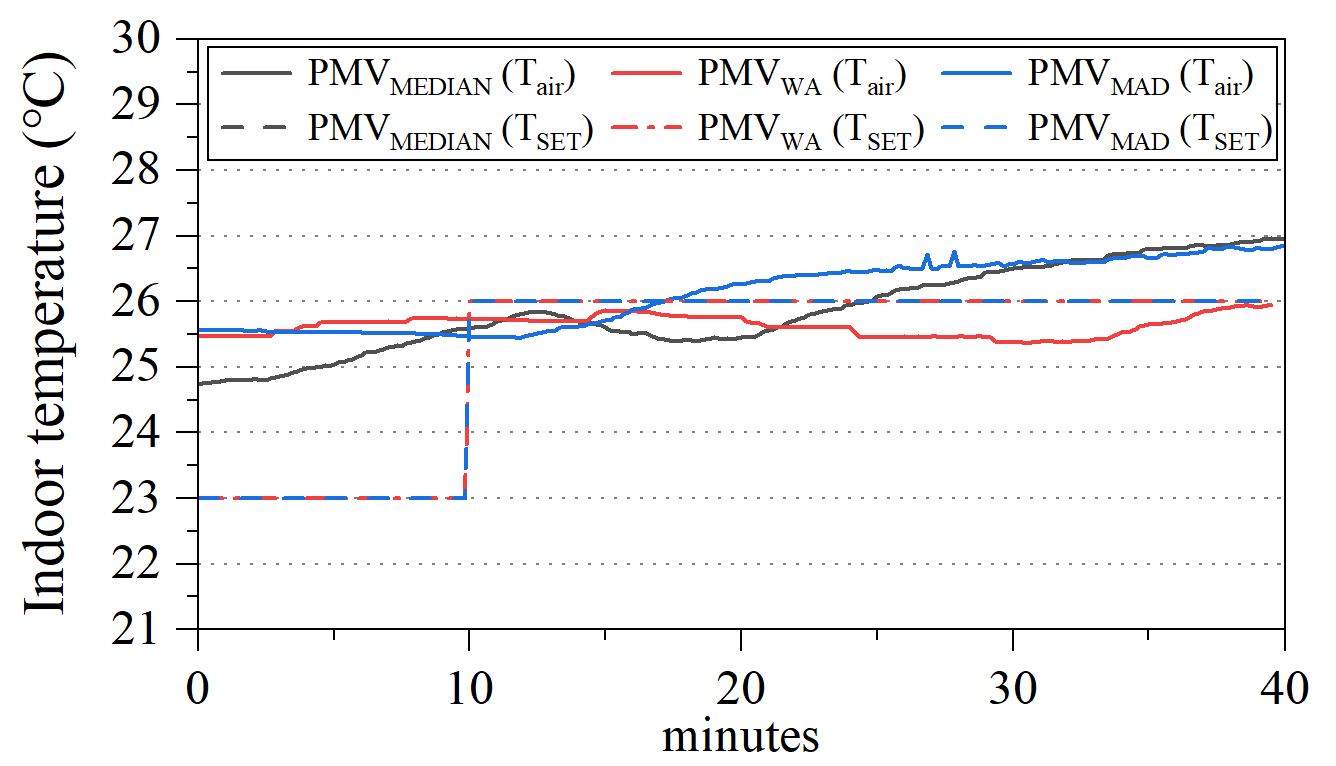}
        \caption{Scenario C}
        \label{fig:results_temp_c}
    \end{subfigure}    
    \hfill
    \begin{subfigure}{0.45\textwidth}
        \centering
        \includegraphics[width=\linewidth]{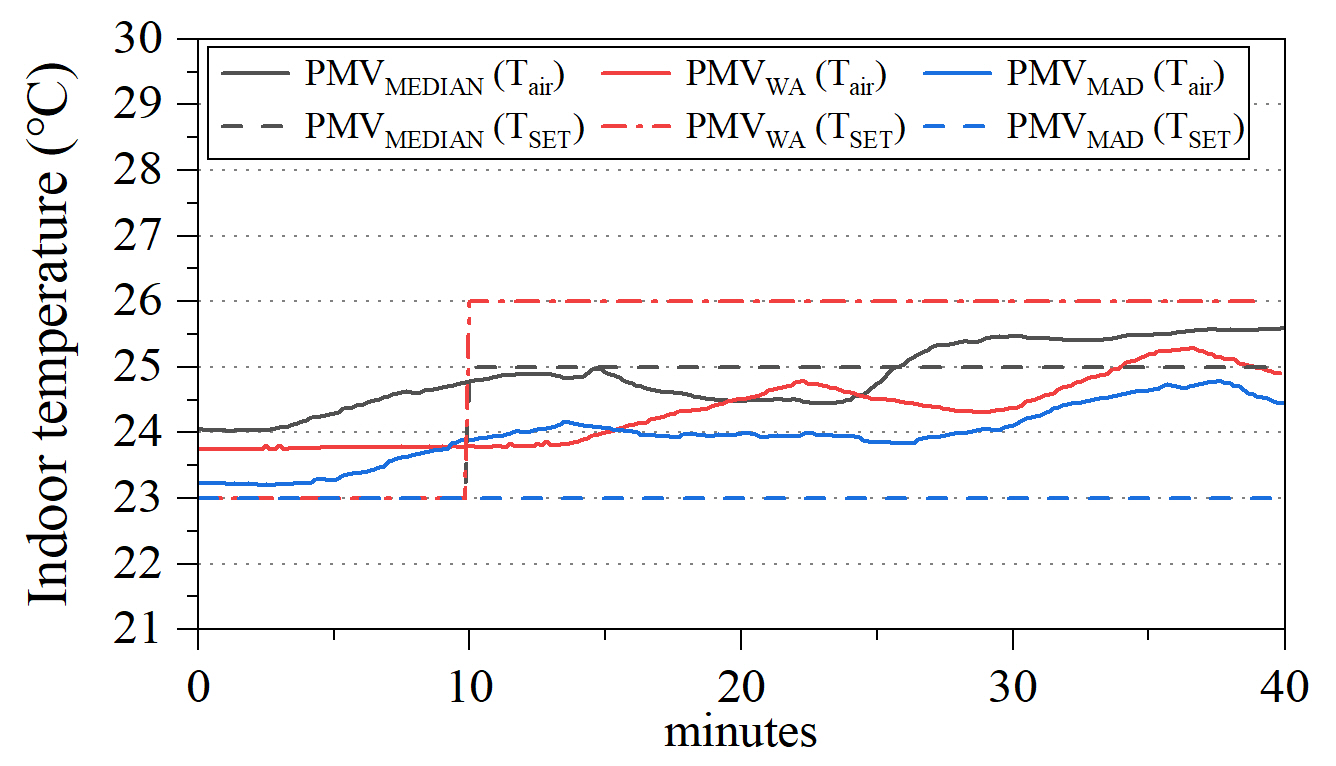}
        \caption{Scenario D}
        \label{fig:results_temp_d}
    \end{subfigure}
    
    \caption{Results of indoor temperature under group PMV-based control}
    \label{fig:results_temp}
\end{figure}

This study implemented indoor thermal control using group PMV values representing the overall thermal comfort of multiple occupants. The corresponding setpoint decisions and indoor temperature changes are shown in Fig.~\ref{fig:results_temp}. Dashed lines indicate the setpoint temperatures (T\textsubscript{SET}) sent to the system at 10 minutes (T\textsubscript{SET1}) and 25 minutes (T\textsubscript{SET2}), while solid lines represent the average of real-time indoor temperature measurements under the control condition. This average was calculated using three sources: (1) two temperature measurements from the indoor sensor modules described in Section~\ref{sec:livinglab}, (2) the 12 boundary measurement points used for temperature reconstruction as shown in Section~\ref{sec:offline}, and (3) the 3 internal validation points introduced in Section~\ref{sec:online} for evaluating reconstruction performance.

As shown in Fig.~\ref{fig:control_process}, the initial control period and the 10-minute preheating at 23\textdegree C were consistently applied. Nevertheless, in some cases, a noticeable increase in indoor temperature was observed during the early phase. This phenomenon was particularly observed in experiments conducted after 2 p.m. (i.e., Scenario A and B under PMV\textsubscript{MEDIAN} and PMV\textsubscript{WA}, and Scenario C). The elevated indoor temperatures compared to the setpoint are not attributed to errors in the control algorithm but rather to the thermal inertia of the building caused by heat accumulation from outdoor conditions.

In all scenarios, the setpoint temperature for PMV\textsubscript{MEDIAN} was determined within the range of 25–26\textdegree C, while PMV\textsubscript{WA} consistently resulted in a fixed setpoint of 26\textdegree C. In contrast, PMV\textsubscript{MAD} produced a wider range of setpoint temperatures, spanning from 22\textdegree C to 26\textdegree C. In Scenarios A, B, and D, PMV\textsubscript{MAD} produced a different setpoint temperature (T\textsubscript{SET}) compared to the other two methods. In contrast, all three methods yielded the same setpoint of 26\textdegree C in Scenario C.

Due to the thermal characteristics of the living lab—where heat tends to accumulate within the experimental space—a time lag was observed between the control input and the actual indoor temperature response, even when the HVAC system functioned as intended. The elevated indoor temperature persisted due to delayed cooling, as observed in Scenario A (PMV\textsubscript{MAD}), where temperatures remained above the setpoint in the final phase. These findings highlight the importance of accounting for thermal inertia in real building control and suggest the need for adaptive control strategies in future applications.

\begin{figure}[htbp]
    \centering
    \begin{subfigure}{0.45\textwidth}
        \centering
        \includegraphics[width=\linewidth]{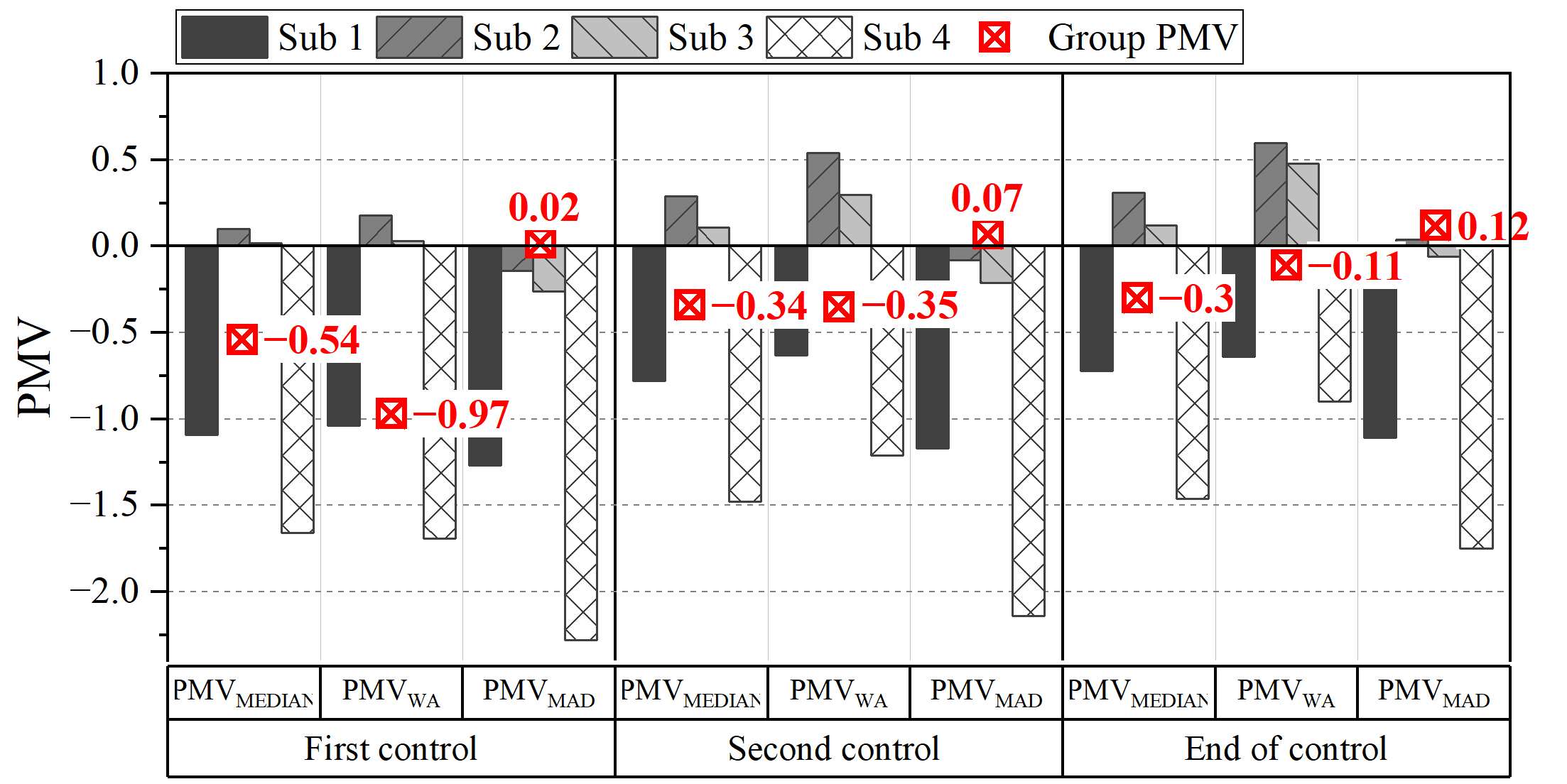}
        \caption{Scenario A}
        \label{fig:results_pmv_a}
    \end{subfigure}
    \hfill
    \begin{subfigure}{0.45\textwidth}
        \centering
        \includegraphics[width=\linewidth]{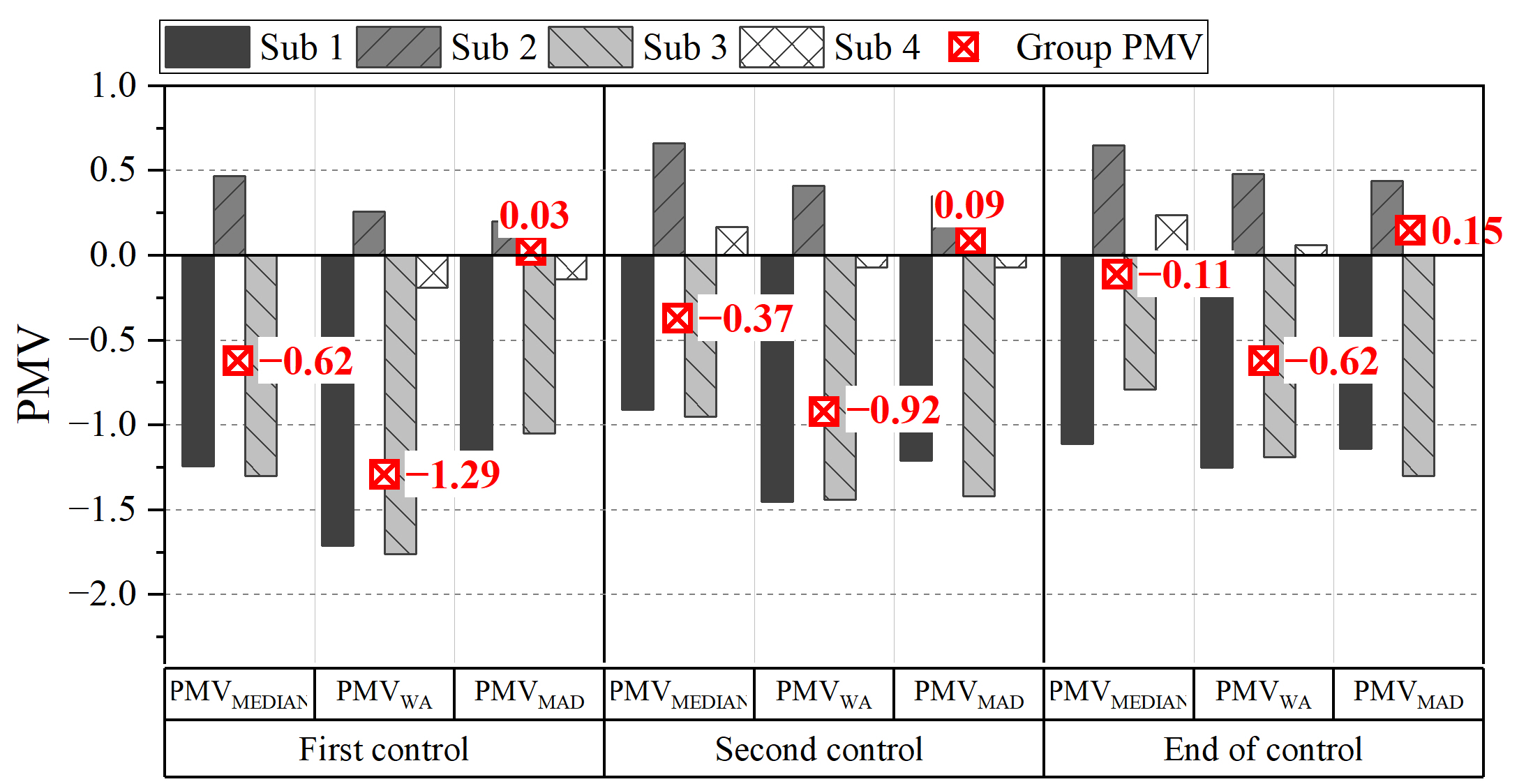}
        \caption{Scenario B}
        \label{fig:results_pmv_b}
    \end{subfigure}
    \vfill
    \begin{subfigure}{0.45\textwidth}
        \centering
        \includegraphics[width=\linewidth]{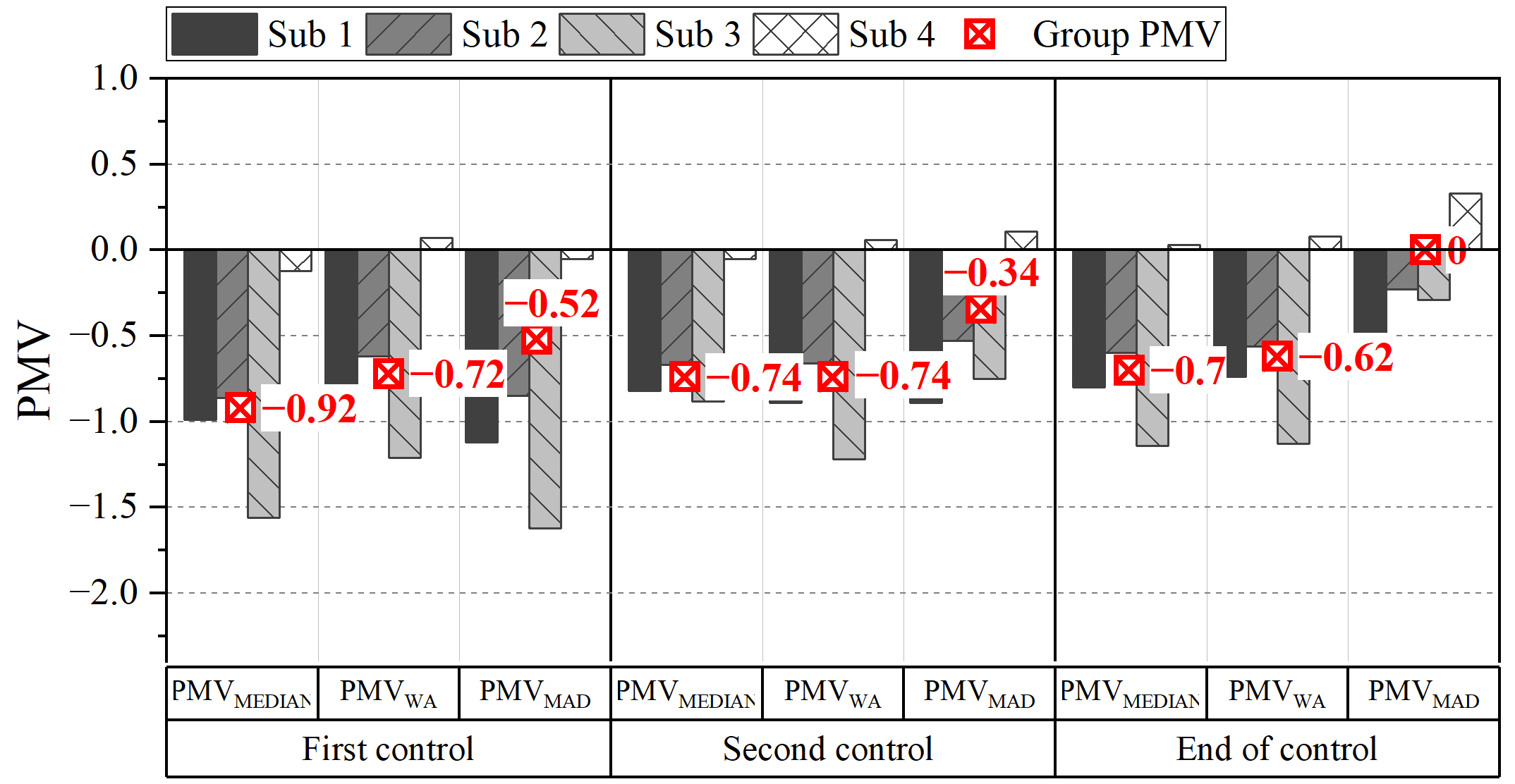}
        \caption{Scenario C}
        \label{fig:results_pmv_c}
    \end{subfigure}    
    \hfill
    \begin{subfigure}{0.45\textwidth}
        \centering
        \includegraphics[width=\linewidth]{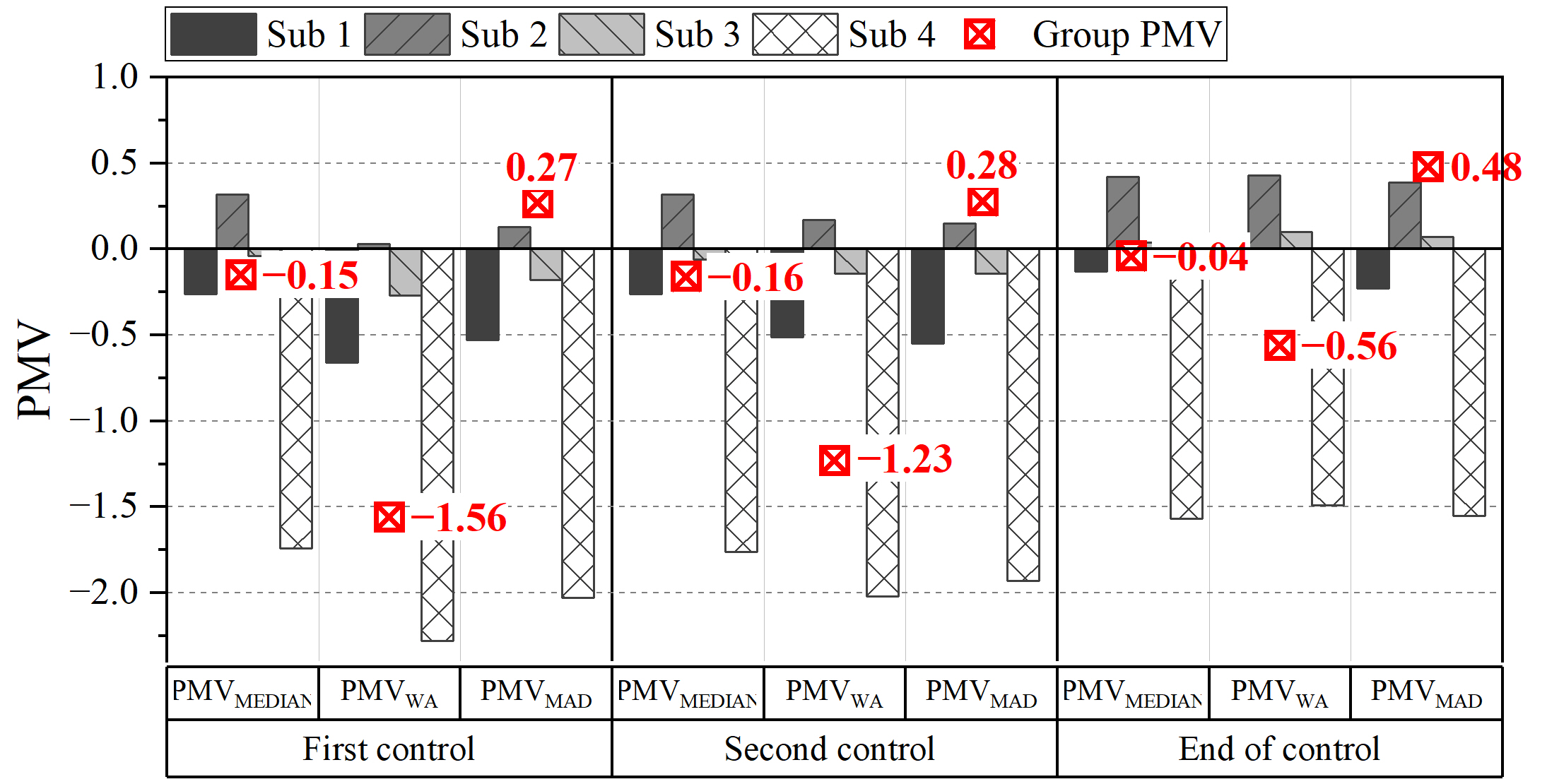}
        \caption{Scenario D}
        \label{fig:results_pmv_d}
    \end{subfigure}
    \caption{Control variable estimation results under different experimental scenarios}
    \label{fig:results_pmv}
\end{figure}

Fig.~\ref{fig:results_pmv} summarizes the control outcomes across all scenarios, presenting both individual PMV values and the corresponding group PMV values computed using three different aggregation methods. These results are shown at three key time points: First control (T\textsubscript{SET1}), Second control (T\textsubscript{SET2}), and End of control. The first control point refers to the moment after 10 minutes of preheating (23\textdegree C) at the beginning of the thermal control period. The second control and end of control points correspond to PMV values recorded after 15 minutes of control at T\textsubscript{SET1} and T\textsubscript{SET2}, respectively.

As discussed in Section~\ref{sec:PMV_variability}, the variation in PMV among participants differed across scenarios. At the first control (T\textsubscript{SET1}), PMV\textsubscript{MAD} generally produced the highest group PMV values, while PMV\textsubscript{WA} consistently resulted in the lowest. The differences between the group PMV methods were pronounced depending on the scenario. This is primarily because PMV\textsubscript{WA} is designed to assign greater weight to participants experiencing thermal discomfort. 
For example, in Scenarios B and D, where participants experienced greater thermal discomfort, PMV\textsubscript{WA} strongly reflected these extremes. 
In contrast, in Scenario C, where individual PMV values showed minimal variation, the differences among all methods were less pronounced. Additionally, greater deviation in individual PMVs led to larger differences between PMV\textsubscript{MAD} and PMV\textsubscript{MEDIAN}, highlighting the influence of participant variability on group-level estimates.

Across all control points, the group PMV values for each method gradually approached the thermal comfort range $(\pm0.5)$ as the experiment progressed. For example, in Scenario A, PMV\textsubscript{WA} improved from -0.97 at the first control (T\textsubscript{SET1}) to -0.11 by the end of the experiment. Similar improvements were observed for PMV\textsubscript{MEDIAN} and PMV\textsubscript{MAD} as shown in Fig. \ref{fig:results_pmv}. These results indicate that adjusting the setpoint temperature using group PMV as a control variable effectively enhanced overall thermal comfort. 

Individual PMV responses varied depending on the group PMV method. PMV\textsubscript{WA} effectively improved comfort for participants with extremely low PMV values—for example, Subject 4 in Scenario A showed improvement in PMV from –1.7 to –0.9. Although this occasionally led to increased discomfort for others (e.g., Subject 2, whose PMV increased from 0.18 to 0.6), it resulted in a tighter distribution around PMV 0 overall. In contrast, PMV\textsubscript{MEDIAN} was less sensitive to outliers but could overlook extreme discomfort in participants with high variability, such as in Scenario D. In Scenario D, Subject 4 showed only limited improvement in comfort under the PMV\textsubscript{MEDIAN} method, with a PMV increase of just 0.19 (from –1.74 to –1.55), whereas PMV\textsubscript{WA} resulted in a considerably larger improvement of 0.76 (from –2.28 to –1.49). PMV\textsubscript{MAD}, which adjusts for deviation, generally produced higher group PMVs when participant differences were large. However, it sometimes led to overheating, especially when indoor temperatures rose excessively. 

These results highlight that different group PMV methods can lead to significantly different control decisions, even under identical occupancy conditions. To further evaluate the effectiveness of these control strategies, participants’ subjective thermal sensations were analyzed across different scenarios.

\begin{figure}[htbp]
    \centering
    \begin{subfigure}{0.45\textwidth}
        \centering
        \includegraphics[width=\linewidth]{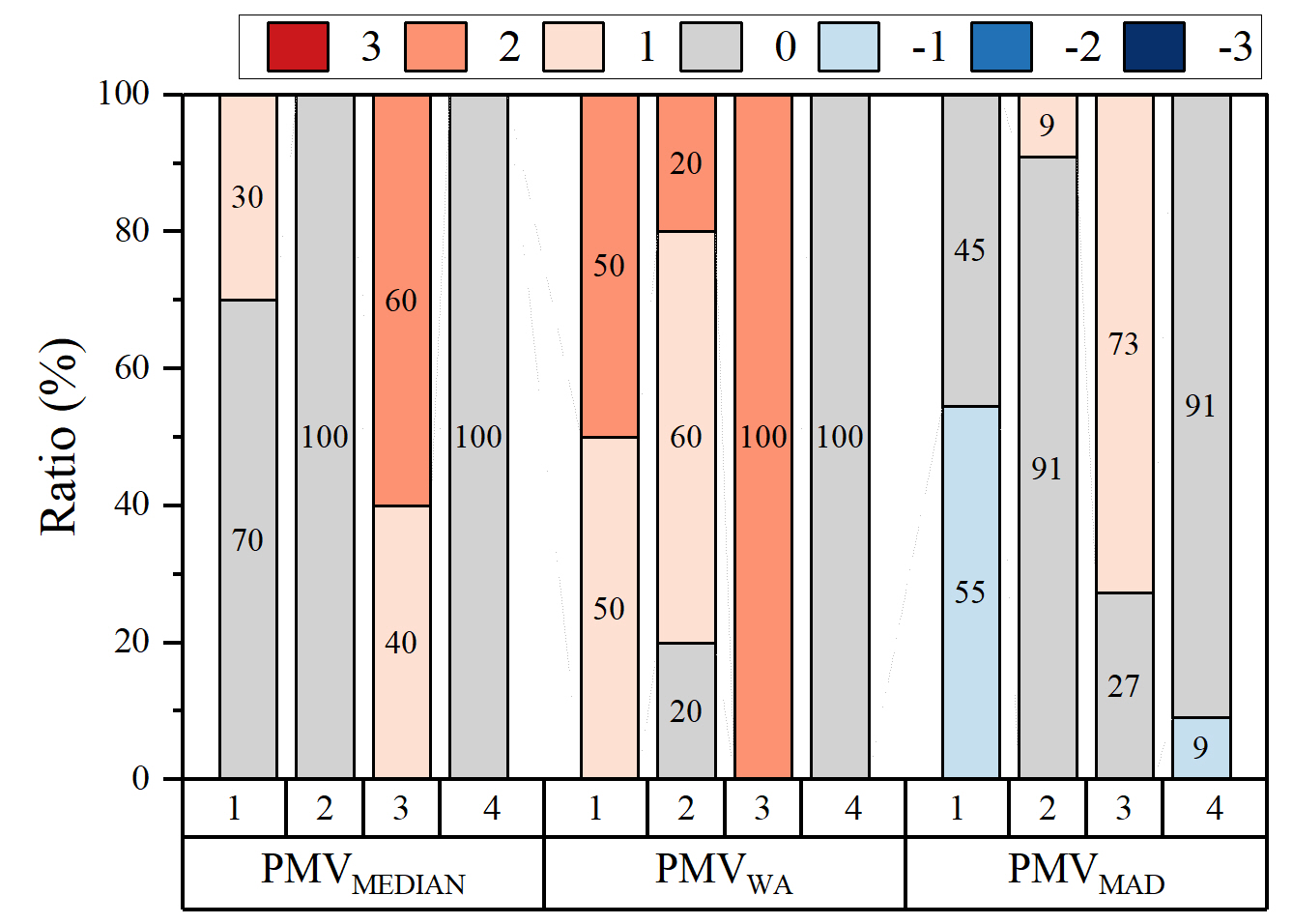}
        \caption{Scenario A}
        \label{fig:results_tsv_a}
    \end{subfigure}
    \hfill
    \begin{subfigure}{0.45\textwidth}
        \centering
        \includegraphics[width=\linewidth]{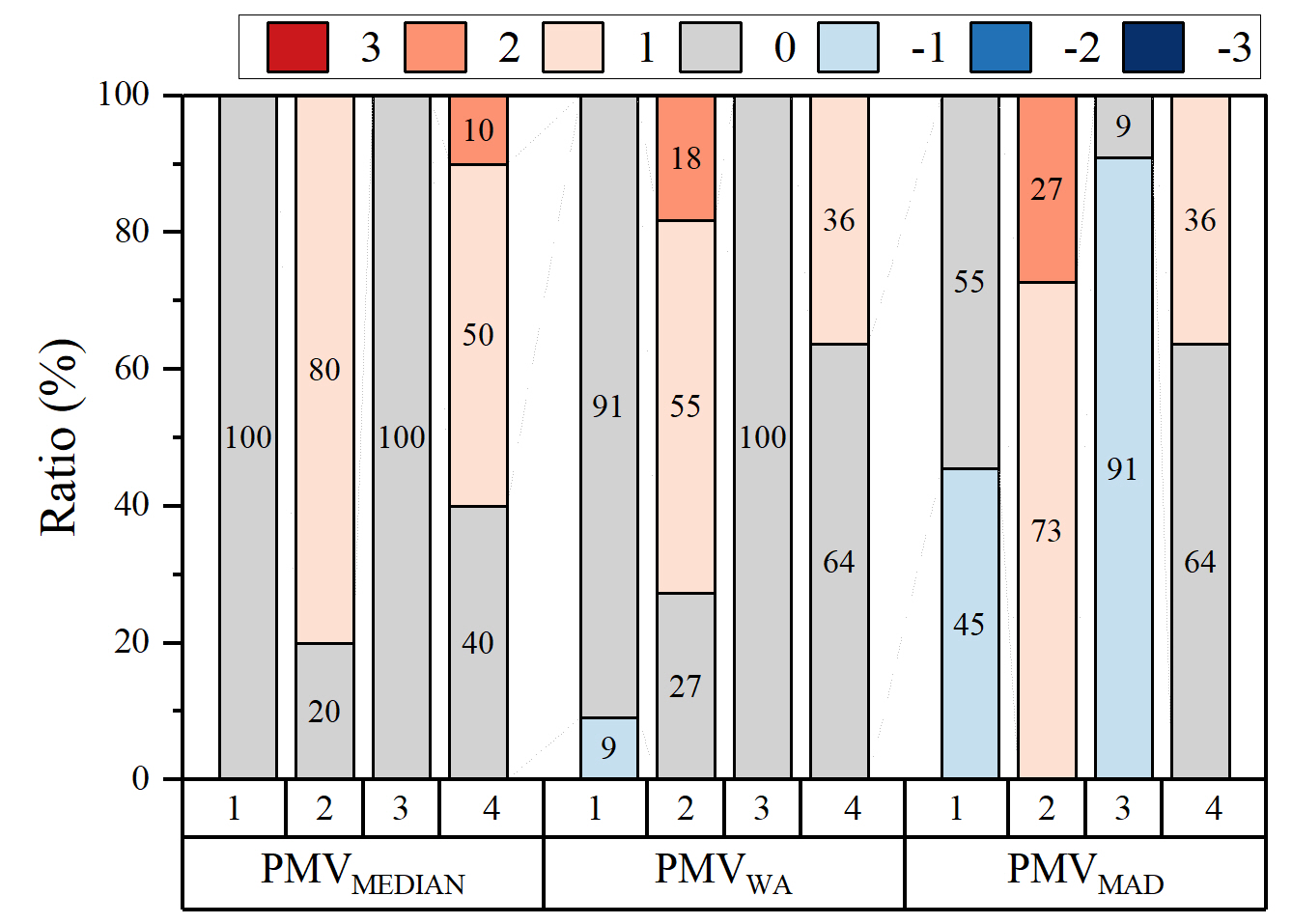}
        \caption{Scenario B}
        \label{fig:results_tsv_b}
    \end{subfigure}
    \vfill
    \begin{subfigure}{0.45\textwidth}
        \centering
        \includegraphics[width=\linewidth]{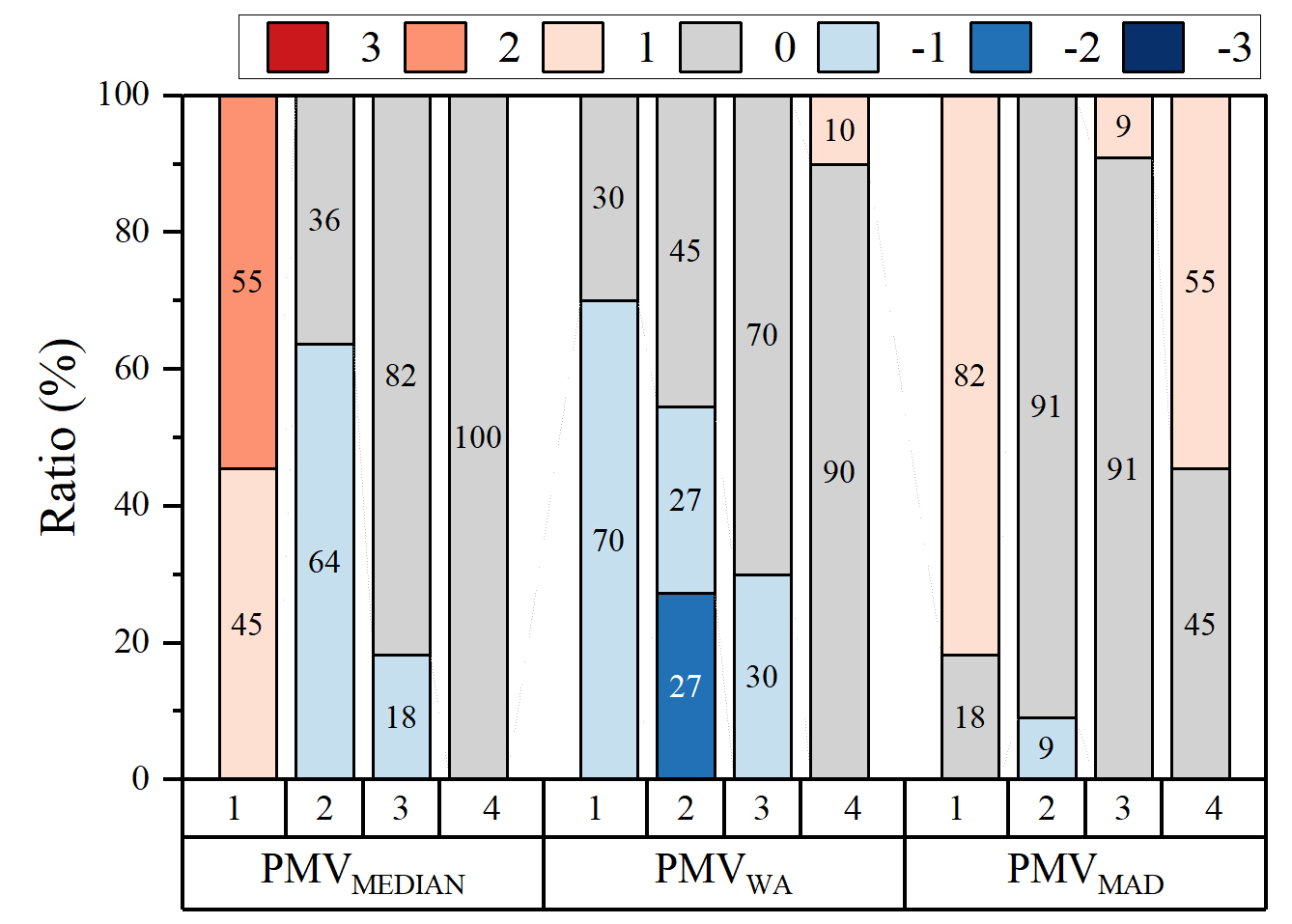}
        \caption{Scenario C}
        \label{fig:results_tsv_c}
    \end{subfigure}    
    \hfill
    \begin{subfigure}{0.45\textwidth}
        \centering
        \includegraphics[width=\linewidth]{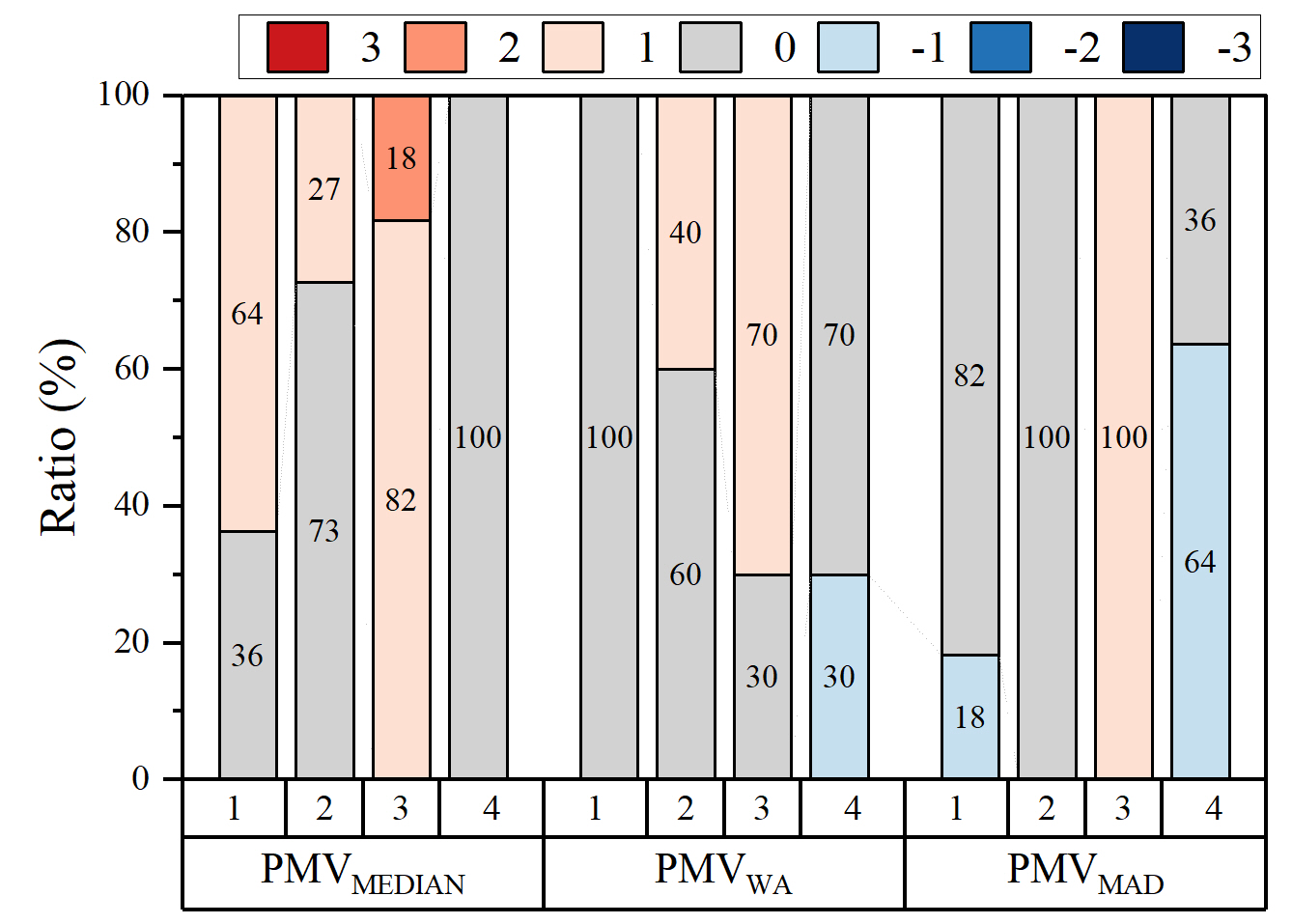}
        \caption{Scenario D}
        \label{fig:results_tsv_d}
    \end{subfigure}
    
    \caption{Distribution of TSV and proportional comfort levels under group PMV-based control methods}
    \label{fig:results_tsv}
\end{figure}

Across 12 experiments, minor issues such as mobile communication errors and missed responses led to an average of 227 TSV entries per participant, totaling 909 responses. Of these, 697 TSV entries collected after control actions were used for analysis. Fig.~\ref{fig:results_tsv} shows the proportion by TSV level.

Perceived TSV results varied by group PMV method, even under the same environmental conditions. Higher indoor temperatures generally led to higher TSV scores, with Scenario A under PMV\textsubscript{WA} showing the highest temperature (26.8\textdegree C) and the greatest discomfort. In this case, most participants—except Subject 4, who had the lowest activity and clothing insulation—reported TSV values above 1.5, indicating a warm sensation. As shown in Fig.~\ref{fig:results_tsv_a}, the proportion of TSV = 2 responses was also highest for PMV\textsubscript{WA}.

In Scenario B, participants with similar personal factors (Subjects 1 and 3) reported high thermal comfort (TSV = 0) under warmer conditions (PMV\textsubscript{MEDIAN}, PMV\textsubscript{WA}), while Subjects 2 and 4 experienced greater discomfort (TSV = 1 or 2) due to contrasting personal factor conditions. The lower indoor temperature under PMV\textsubscript{MAD} resulted in a higher proportion of TSV = -1 responses, indicating that all subjects experienced discomfort.
In Scenarios C and D, despite comparable average indoor temperatures across the three control methods, variations in TSVs were observed. These differences are attributed to localized environmental factors, such as late-stage temperature increases and occupant positioning within the space. For example, in Scenario D, Subject 3 who is positioned near the air outlet, reported higher thermal sensation than Subject 2, despite identical personal conditions. These findings emphasize that spatial factors like airflow and occupant position significantly affect perceived comfort.

To assess the control performance of each group PMV method, participant TSV responses were analyzed using two comfort ranges, as shown in Table~\ref{tab:results_comfort_range}: complete neutrality (TSV = 0) and broader comfort with low thermal discomfort (-1 ≤ TSV ≤ +1). 

Based on the analysis, the proportion of participants reporting complete neutrality (TSV = 0) followed the order PMV\textsubscript{MEDIAN} (62.2\%) > PMV\textsubscript{WA} (58.0\%) > PMV\textsubscript{MAD} (52.8\%). None of the methods exceeded 70\% on average, highlighting limitations in achieving universal thermal comfort across all occupants. Among the three approaches, the PMV\textsubscript{MEDIAN} method exhibited the most consistent performance across varying personal factor conditions. In contrast, PMV\textsubscript{WA} and PMV\textsubscript{MAD} showed lower TSV = 0 proportions in scenarios with high inter-individual variation, indicating reduced effectiveness in settings with diverse occupant profiles.

Analysis of the broader comfort range (–1 ≤ TSV ≤ +1) revealed that all three group PMV estimation methods achieved average thermal comfort rates exceeding 86.5\%, with PMV\textsubscript{MAD} showing the highest performance at 98.4\%. In comparison to the PMV values presented in Fig.~\ref{fig:results_pmv}, the actual TSV responses exhibited a tendency toward higher perceived thermal sensation. For instance, PMV values between –2 and –1 often corresponded to TSV = –1, while PMV > +0.5 frequently aligned with TSV = +2. This indicates a positive perceptual bias, where subjective thermal sensation is higher than objective PMV estimates. The PMV\textsubscript{MAD} method helped reduce this discrepancy by accounting for distributional deviations. While PMV\textsubscript{MEDIAN} was less sensitive to individual extremes, it provided consistent control across scenarios. In contrast, PMV\textsubscript{WA} effectively addressed extreme discomfort but showed greater variability, particularly under various occupant conditions.

On average, PMV\textsubscript{MAD} demonstrated the most effective subjective thermal performance across various scenarios; however, these results also highlight the trade-offs among different group PMV estimation methods. PMV\textsubscript{MEDIAN} offers stable and balanced performance across varied conditions, making it well-suited for general comfort control. PMV\textsubscript{MAD} is effective in mitigating extreme discomfort by incorporating variability, though it may lower the proportion of neutral responses. PMV\textsubscript{WA} enhances comfort for highly dissatisfied occupants but exhibits inconsistent results at the group level. Therefore, the selection of an appropriate group PMV method should be based on the specific context of application, including the space’s function, occupant characteristics, and control objectives.

\begin{table}[htbp]
\caption{Occupants thermal satisfaction ratio by thermal comfort range}
\label{tab:results_comfort_range}
\begin{adjustbox}{width=\textwidth}
\renewcommand{\arraystretch}{1.3}
\begin{tabular}{lllllll}
\hline
Comfort range    & \multicolumn{3}{c}{TSV   = 0}                                             & \multicolumn{3}{c}{-1 ≤ TSV ≤ +1}   \\ 
Methods          & PMV\textsubscript{MEDIAN} & PMV\textsubscript{WA} & PMV\textsubscript{MAD} & PMV\textsubscript{MEDIAN} & PMV\textsubscript{WA} & PMV\textsubscript{MAD} \\ \hline
Scenario A     & 67.5\%  & 30.0\%  & 63.6\%  & 85.0\%  & 57.5\%  & 100\% \\
Scenario B     & 65.0\%  & 70.5\%  & 31.8\%  & 97.5\%  & 95.5\%  & 93.2\%  \\
Scenario C     & 54.5\%  & 58.9\%  & 61.4\%  & 86.4\%  & 93.2\%  & 100\%  \\
Scenario D     & 52.3\%  & 65.0\%  & 54.5\%  & 95.5\%  & 100\%  & 100\% \\ 
Total mean     & 59.8\%  & 56.1\%  & 52.8\%  & 91.1\%  & 86.5\%  & 98.3\%  \\ \hline
\end{tabular}
\end{adjustbox}
\end{table}

\section{Conclusion}\label{sec:conclusion}
\subsection{Summary of Key Findings}
This study proposes a PMV estimation approach that utilizes an indoor temperature reconstruction algorithm based on Gappy POD to account for occupant location. Spatial variations in PMV across multi-occupant conditions were quantitatively analyzed, and a group PMV-based control framework was applied to evaluate thermal control performance. 
The Gappy POD method reconstructs real-time indoor temperature distributions using a limited number of sensor measurements. This approach enables fast and efficient reconstruction by leveraging a pre-trained POD basis. Also, it supports accurate estimation of spatially varying PMV values. Experimental results from a living lab indicate that the average relative error in temperature reconstruction remained below 3\%, demonstrating the method’s feasibility and reliability.

Thermal comfort analysis based on the reconstructed indoor temperature data in the living lab revealed that spatial variations in PMV across the space can reach up to 1.26 scale units. By integrating reconstructed environmental fields with occupant-specific personal factors, accurate location-based PMV values were estimated and applied to group PMV-based control strategies. The results demonstrated that each group PMV method produced distinct thermal comfort outcomes depending on the occupancy scenario. When thermal satisfaction was assessed by the proportion of occupants within the comfort range (TSV $\leq \pm 1$), the MAD-based method which accounts for distributional variability proved more effective in enhancing thermal comfort during the heating season. 

In summary, the findings of this study highlight the feasibility of real-time thermal comfort estimation and control based on occupant location, even within spatially heterogeneous indoor environments. In addition, the results underscore the importance of selecting appropriate group PMV metrics in accordance with occupant diversity and system objectives. Furthermore, the study emphasizes the necessity of thermal control strategies that integrate both spatial and individual variability to ensure optimal comfort in shared environments. 

\subsection{Limitations and Future research directions}
Despite the contributions of this study, several technical limitations were identified, highlighting important directions for future research. First, regarding the data reconstruction algorithm, the effect of sensor placement on reconstruction performance was not investigated. This is because sensor placement was assumed, in which sensors were evenly distributed along each boundary to reflect practical constraints related to building design and occupant convenience. In addition, the number of sensors used in the living lab (12 sensors in a 3.2\,m $\times$ 9.45\,m space) is relatively large. Further research is needed to achieve high reconstruction accuracy with fewer sensors. While Gappy POD is effective for reconstructing quantities with minimal advection effects, its performance may degrade in scenarios where strong internal airflow introduces dominant advection dynamics.

To overcome these limitations, several future directions are planned. Specifically, the influence of sensor placement on reconstruction accuracy will be quantitatively evaluated. Optimization algorithms such as the Discrete Empirical Interpolation Method (DEIM) \cite{chaturantabut2010nonlinear, drmac2016new, drmac2018discrete} and S-OPT \cite{shin2016nonadaptive, lauzon2022s} will be employed for sensor placement, and their reconstruction performance will be systematically compared. DEIM is a greedy sampling algorithm that selects interpolation points to minimize the upper bound of reconstruction error. In addition to the original DEIM formulation, an oversampled variant \cite{carlberg2011efficient, lauzon2022s} is considered, where the number of sampling points is allowed to exceed the number of POD modes. S-OPT is a quasi-optimal sampling method that seeks to maximize the orthogonality of the sampled POD basis vectors and the determinant of the corresponding Gram matrix constructed from them, thereby enhancing numerical stability and reducing reconstruction error. 

Furthermore, the application of Gappy Autoencoder (Gappy AE) \cite{KIM2024116978}, a nonlinear manifold-based reconstruction algorithm, is planned to be explored. Gappy AE offers the potential to achieve high reconstruction accuracy with fewer sensors and is particularly suitable for reconstructing environmental quantities governed by advection-dominant dynamics. In particular, Gappy AE is expected to perform well in reconstructing airflow velocity fields, which typically exhibit strong advection and complex flow structures.

The control experiment also had several limitations. Due to the structural characteristics of the experimental space and the aging HVAC system, precise temperature control was constrained, and the setpoint range was limited to 22–26\textdegree C. Additionally, some data loss occurred due to sensor noise, communication delays, and storage errors. However, to minimize the impact on control performance, data collected during the two minutes preceding each control point were preprocessed using mean and most frequent values. Furthermore, the experiment was conducted under predefined and controlled scenarios, which did not fully reflect the dynamic and unpredictable nature of real-world occupancy. To enhance applicability in actual environments, the integration of real-time occupant detection technologies is necessary, in addition to the environmental reconstruction algorithm and the PF model proposed in this study.

This study was conducted during the heating season with a limited number of participants. Therefore, future research should involve long-term data collection encompassing a wider range of seasonal conditions and more diverse occupancy patterns. In particular, systematically gathering thermal comfort responses under varying HVAC operating modes and occupancy scenarios will facilitate the development of more robust and generalizable models for real-world applications.
From the perspective of occupant data, it is necessary to incorporate real-time occupant detection and location-tracking capabilities to extend the system's applicability to large-scale smart building environments. Privacy concerns associated with image-based AI models must also be carefully addressed to ensure feasibility in practical deployment.
Lastly, a method for determining a control variable that represents group thermal comfort and dynamically integrates real-time information—such as occupancy status, number of occupants, metabolic rate, and clothing insulation—is essential. Such capabilities would advance the system from scenario-based control to an adaptive, user-centered thermal management framework capable of responding effectively to real-world environmental changes.

\section*{Acknowledgments}
This work was supported by the National Research Foundation of Korea(NRF) grant funded by the Korea government(MSIT) (RS-2023-00217322 and RS-2023-00276529) and  the Ministry of Trade, Industry, and Energy and the Korea Evaluation Institute of Industrial Technology research grant (20012462). 

\bibliographystyle{plain}
\bibliography{ref}  

\end{document}